\documentclass[usenatbib]{mnras}
\pdfoutput=1
\usepackage{mathptmx}
\usepackage{txfonts}

\usepackage[T1]{fontenc}
\usepackage{ae,aecompl}

\usepackage{graphicx}
\usepackage[space]{grffile}
\usepackage[utf8]{inputenc}
\usepackage[english]{babel}

\newcommand{\Rs}{\ensuremath{R_{\odot}}}
\newcommand{\Ms}{\ensuremath{M_{\odot}}}
\newcommand{\Zs}{\ensuremath{Z_{\odot}}}
\newcommand{\eg}{{\it e.g.}}
\newcommand{\cf}{{\it c.f.~}}
\newcommand{\ie}{{\it i.e.}}

\newcommand{\beq}{\begin{equation}}
\newcommand{\eeq}{\end{equation}}

\newcommand{\kmps}{\ensuremath{{\rm~km~s}^{-1}}}

\newcommand{\nbseven}{{\tt NBODY7~}}
\newcommand{\nbsix}{{\tt NBODY6 }}
\newcommand{\nbpp}{{\tt NBODY6++ }}
\newcommand{\bse}{{\tt BSE }}
\newcommand{\sse}{{\tt SSE }}

\newcommand{\ptsone}{{\tt pts1}}
\newcommand{\ptstwo}{{\tt pts2}}
\newcommand{\ptsthree}{{\tt pts3}}

\newcommand{\nbh}{\ensuremath{N_{\rm BH}}}

\newcommand{\mrem}{\ensuremath{M_{\rm rem}}}

\newcommand{\fmrg}{\ensuremath{f_{\rm mrg}}}
\newcommand{\ftz}{\ensuremath{f_{\rm TZ}}}
\newcommand{\mmax}{\ensuremath{m_{\rm max}}}

\title[Massive black hole formation at high metallicity]
{On the formation of exotic, massive, stellar-remnant black holes at solar and sub-solar
metallicities through evolution
of massive binaries}

\author[S. Banerjee]{
Sambaran Banerjee$^{1,2}$\thanks{E-mail: sambaran@astro.uni-bonn.de (SB)}
\\
$^{1}$Helmholtz-Instituts f\"ur Strahlen- und Kernphysik (HISKP),
Nussallee 14-16, D-53115 Bonn, Germany\\
$^{2}$Argelander-Institut f\"ur Astronomie (AIfA),
Auf dem H\"ugel 71, D-53121, Bonn, Germany
}

\pubyear{2020}

\begin{document}
\label{firstpage}
\pagerange{\pageref{firstpage}--\pageref{lastpage}} 
\maketitle

\begin{abstract}
The recent inference of a $70\Ms$ black hole (BH) in the Galactic, detached binary LB-1
has sparked cross-disciplinary debate since a stellar remnant of such large
mass is well above what can be expected from stellar-evolutionary theory,
especially in an enriched environment like that of the Milky Way. This
study focusses on the possibilities of formation of extraordinarily massive  
BHs at solar and globular cluster (GC)-like metallicities via evolution of massive
stellar binaries. A population-synthesis
approach is followed utilizing the recently-updated {\tt BSE} program.
BHs in the mass range of $50\Ms-80\Ms$ could be formed at the solar metallicity only
if a large fraction, $\gtrsim70$\%, of matter is allowed to accrete onto a low-mass
BH, in a BH-star merger product (a ``black hole Thorne-Zytkow object''; BH-TZO).
Their counterparts at GC-like metallicities can reach $100\Ms$. Although
post-accretion BHs can, generally, be expected to be of high spin parameter,
they can potentially be of low spin in the case of a BH-TZO.
This spin aspect remains speculative in this work and deserves detailed hydrodynamic studies.
\end{abstract}


\section{Introduction}\label{intro}

The recent discovery of a detached binary in our Galaxy's field, comprising of a $68_{-13}^{+11}\Ms$
black hole (hereafter BH) and a B-type star of $8.2_{-1.2}^{+0.9}\Ms$ in a $78.9$-day, near-circular
(eccentricity $e=0.03\pm0.01$) orbit, as inferred by \citet{Liu_2019} through radial-velocity measurements,
has become a focal topic across disciplines in astronomy. While other detached BH-star binaries
have also been discovered over the last couple of years in Galactic globular clusters (hereafter GC) and
in the field \citep{Giesers_2018,Giesers_2019,Thompson_2019}, the above binary, nicknamed LB-1,
is much of an exception due to its BH member's $\approx70\Ms$ mass.
As opposed to this, the BH members of all the other, to-date known Galactic BH-star binaries are estimated
to be of $<10\Ms$ (the unseen member in the APOGEE binary of \citealt{Thompson_2019} can, in fact,
be an exceptionally-massive neutron star; hereafter NS).

Note that LIGO-Virgo observations of gravitational waves (hereafter GW) from binary black hole (hereafter BBH)
mergers have, so far, identified up to $\approx50\Ms$ BH \citep{Abbott_GWTC1}, which,
after taking into account the measurement uncertainties,
is consistent with the $\approx40\Ms$ upper limit expected for stellar-remnant BHs due to
pulsation pair-instability supernova \citep{Langer_2007,Woosley_2017}. From that point of view,
a $70\Ms$ BH is indeed intriguing but can potentially be explained as a BBH merger
product (\eg, \citealt{Rodriguez_2016,Banerjee_2017,DiCarlo_2019}) or as the remnant of a star-star
merger (\eg, \citealt{Banerjee_2020,Spera_2019}). However,
such merger-based scenarios would still require sub-solar metallicities,
given that at the solar metallicity
single-stellar evolution would yield up to $\approx15\Ms$ BH even for a very large
zero age main sequence (hereafter MS; ZAMS) mass \citep{Hurley_2000,Belczynski_2010,Banerjee_2020},
as per the current understanding of stellar wind mass loss. Therefore, the finding of a $70\Ms$ BH
in the Milky Way's field clearly presents challenges to our current understanding
of stellar evolution. Note that
LB-1's observed spectral variability is, as well, susceptible to alternative
interpretations that would point to a much lower mass of its invisible member
\citep{ElBadry_2019,AbdulMasih_2019}.

In this context, this work follows a population synthesis-based approach to investigate the conditions,
however exotic they may be, under which BHs of $50\Ms-80\Ms$ can form in solar-metallicity
(or near-solar-metallicity) environments. In particular, observationally-motivated
massive binary populations are evolved, using an updated version of the $\bse$ binary-evolution
program, to study their remnant outcomes. The possibilities
and mechanisms for forming such massive BHs at low metallicities are also explored. This
paper is organized as follows: Sec.~\ref{method}
describes the $\bse$ program, the model massive-binary population,
and the adopted physical conditions. Sec.~\ref{res} describes
the results. Sec.~\ref{discuss} discusses the caveats, outlooks, and future prospects.

\begin{figure*}
\includegraphics[width=6.0cm,angle=0]{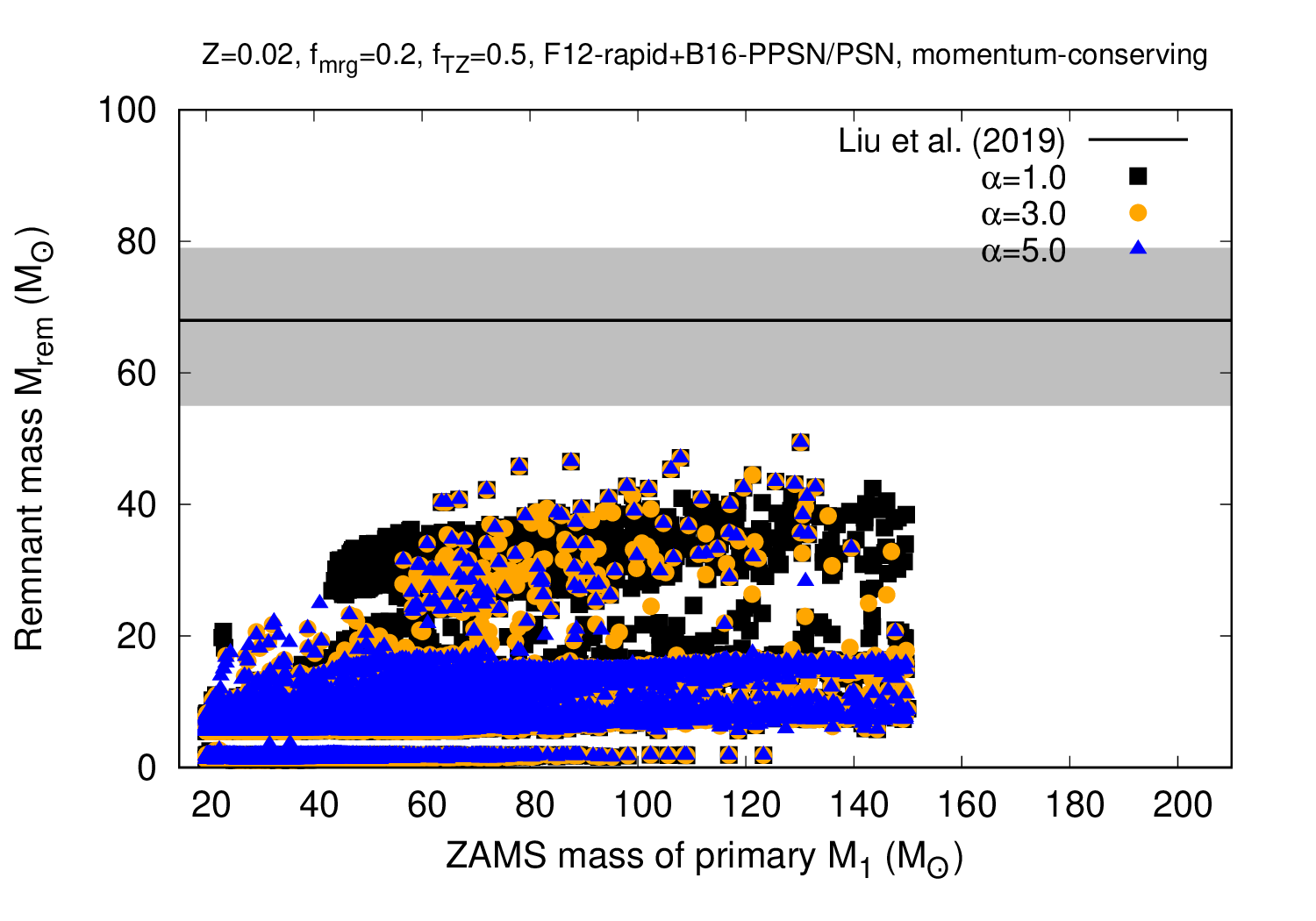}
\hspace{-0.5 cm}
\includegraphics[width=6.0cm,angle=0]{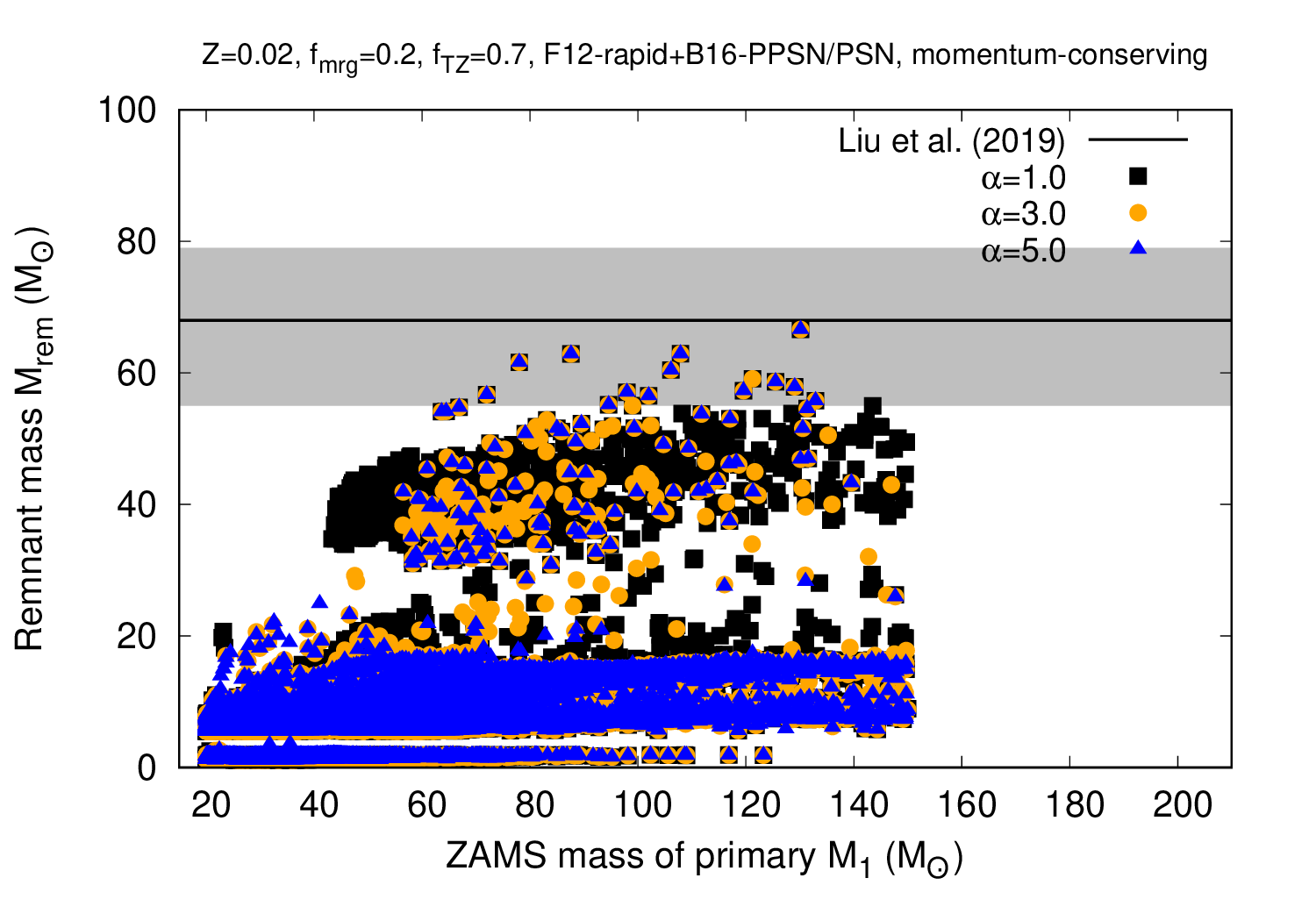}
\hspace{-0.5 cm}
\includegraphics[width=6.0cm,angle=0]{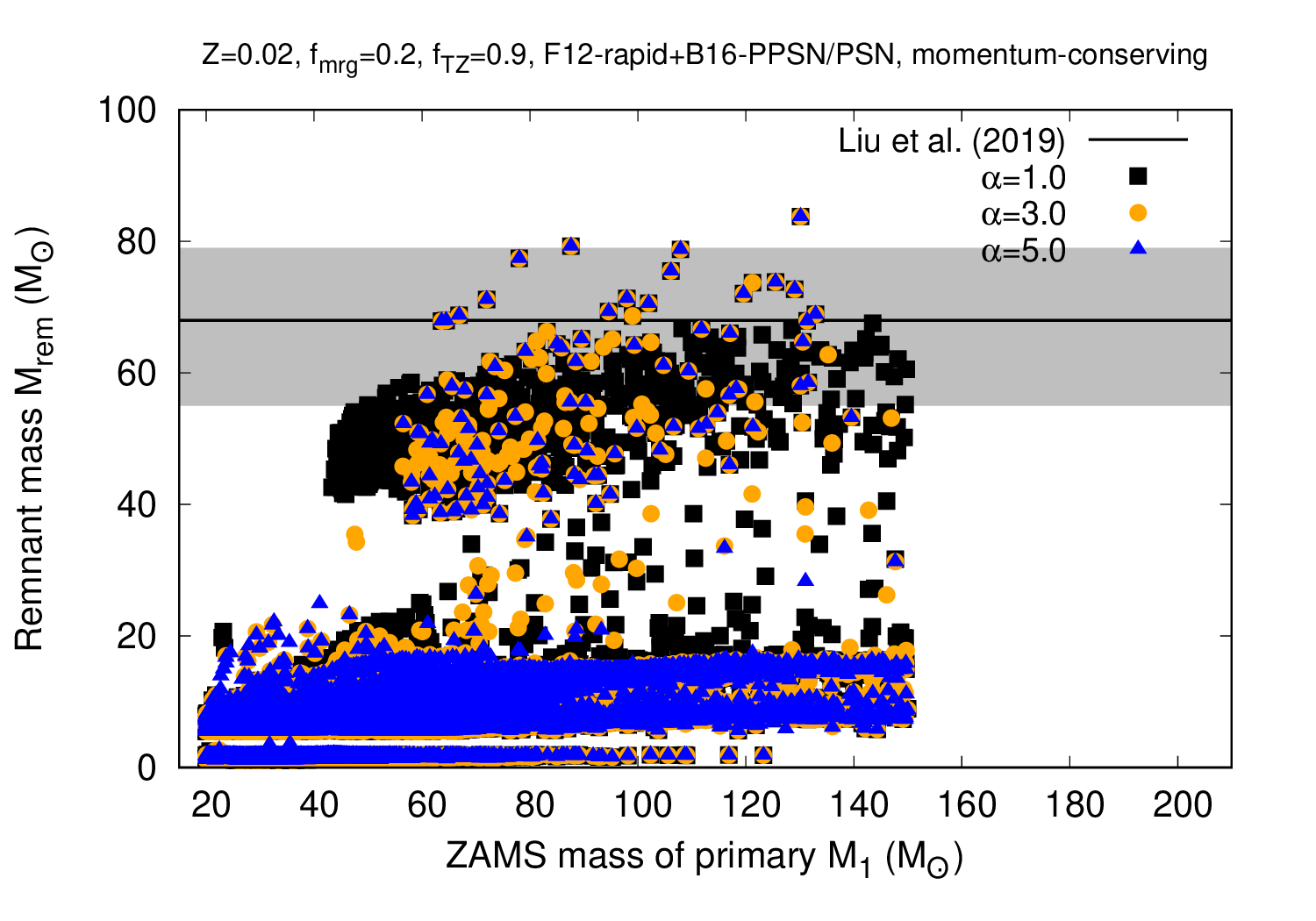}\\
\includegraphics[width=6.0cm,angle=0]{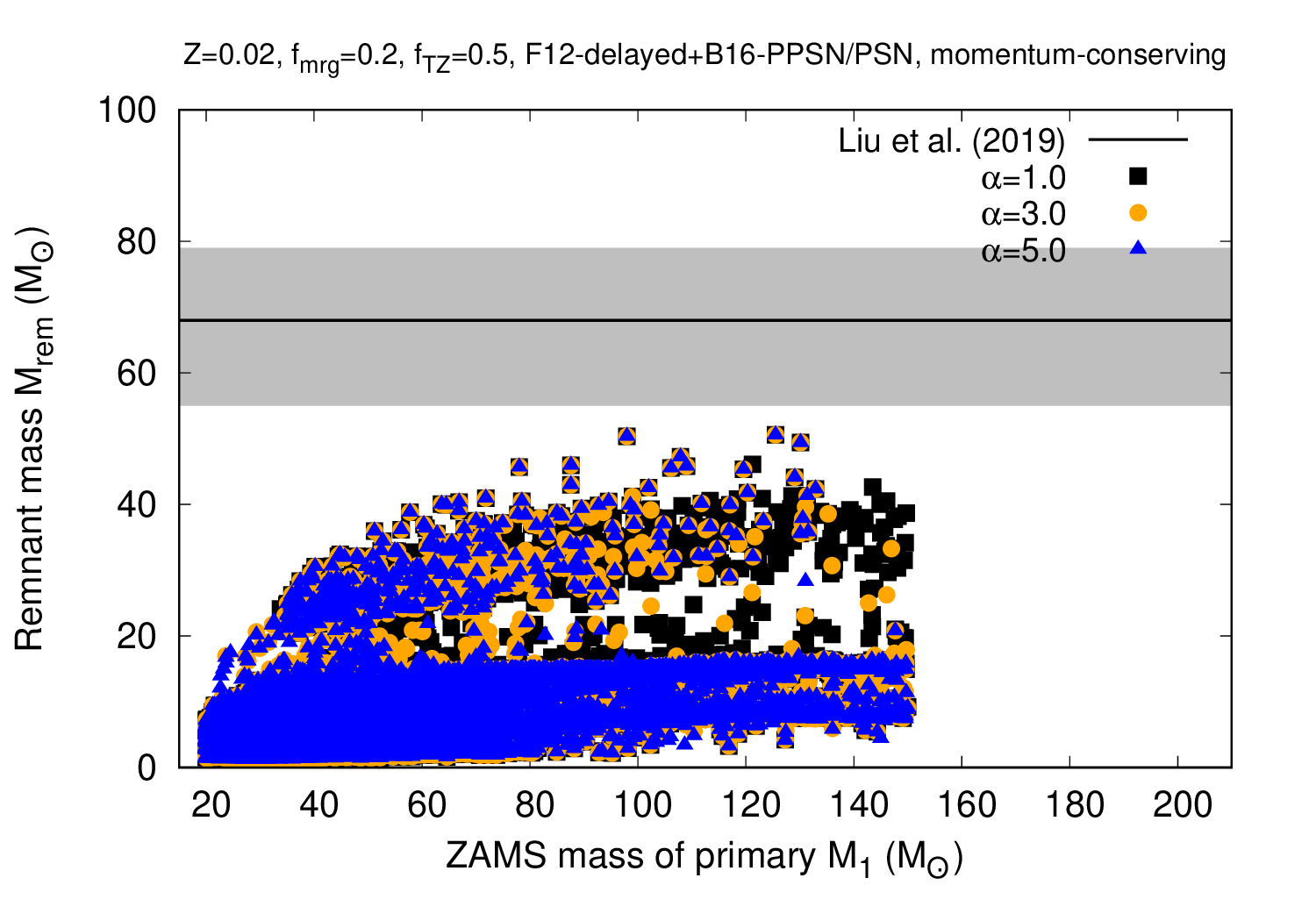}
\hspace{-0.5 cm}
\includegraphics[width=6.0cm,angle=0]{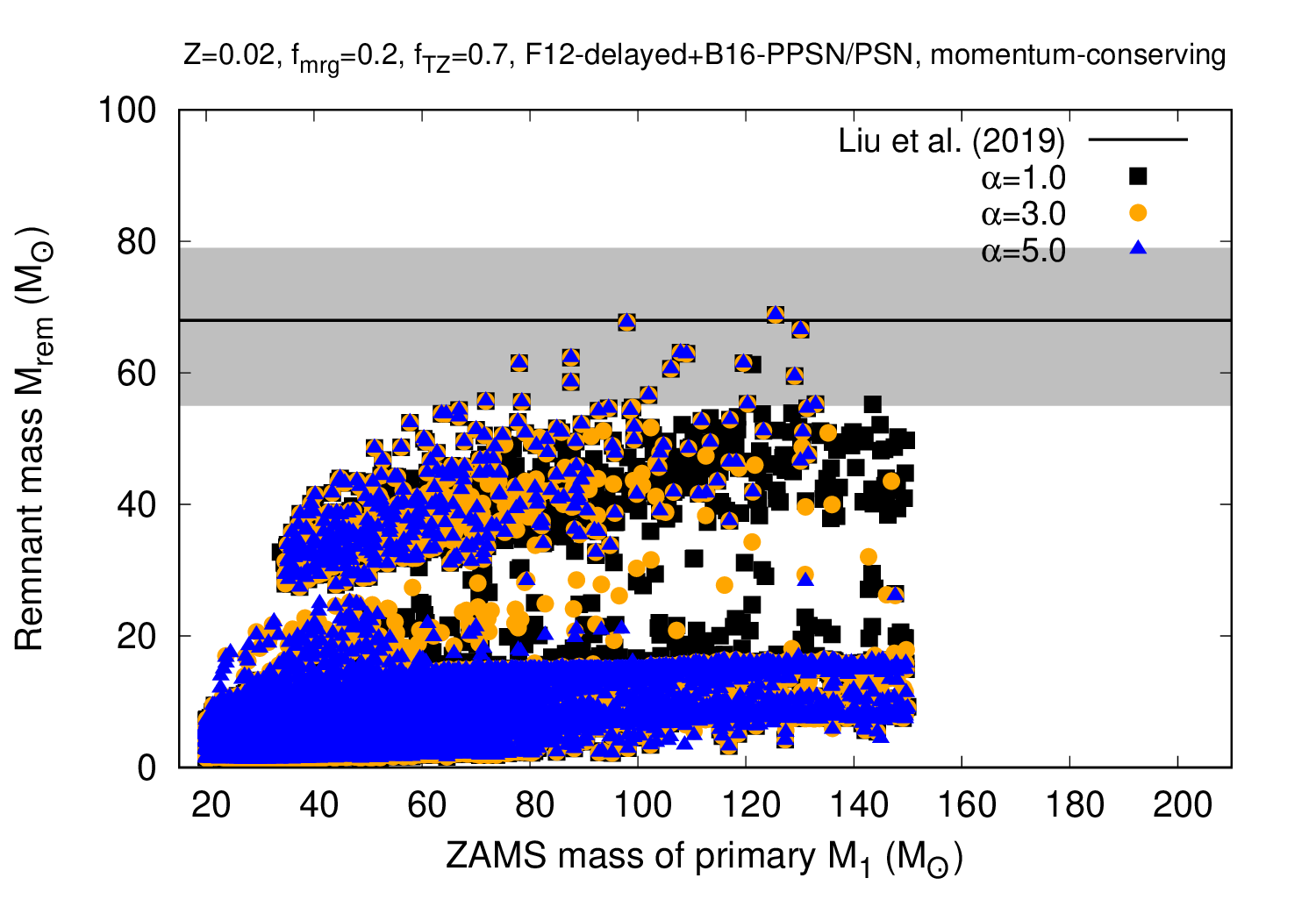}
\hspace{-0.5 cm}
\includegraphics[width=6.0cm,angle=0]{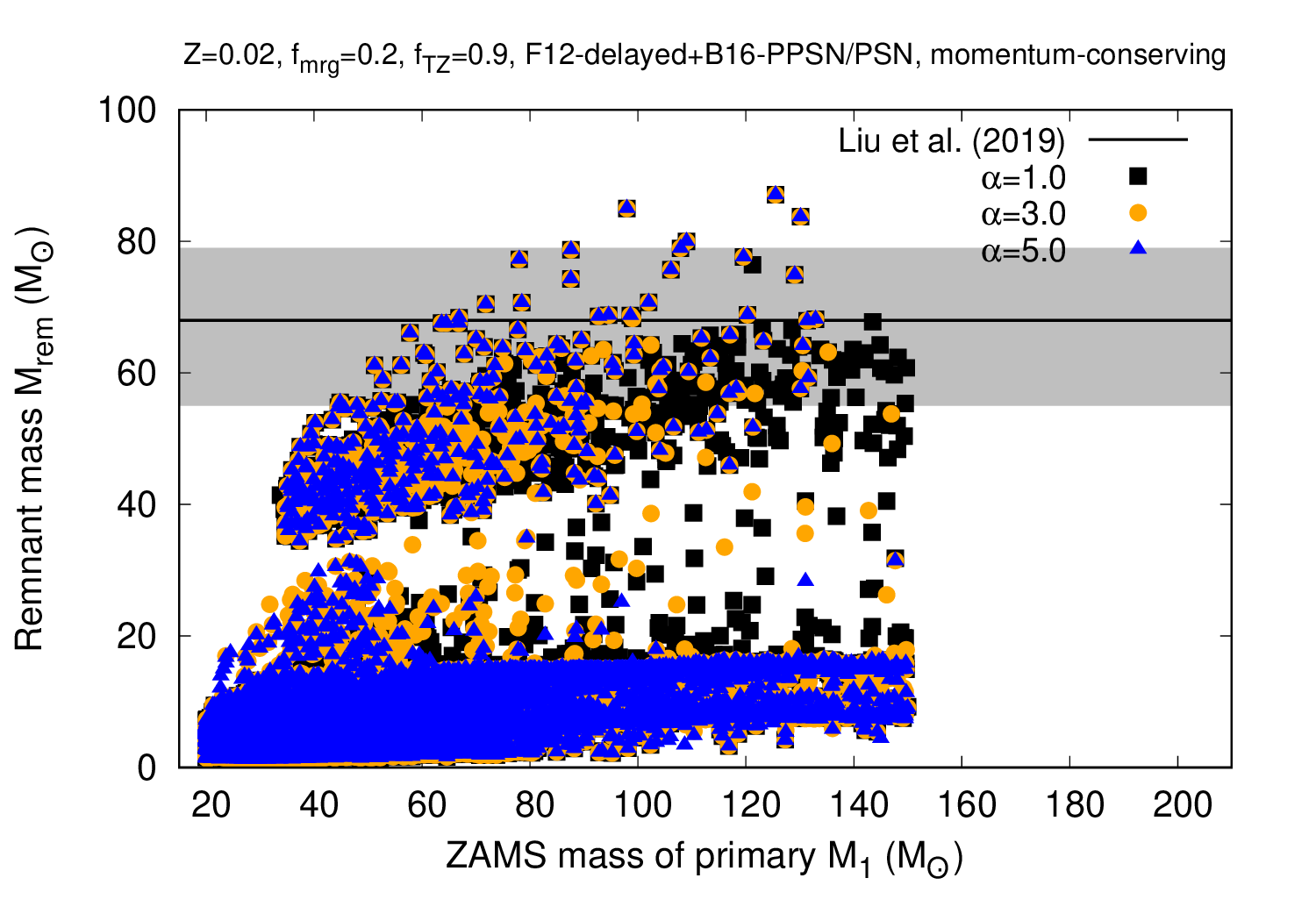}\\
\includegraphics[width=6.0cm,angle=0]{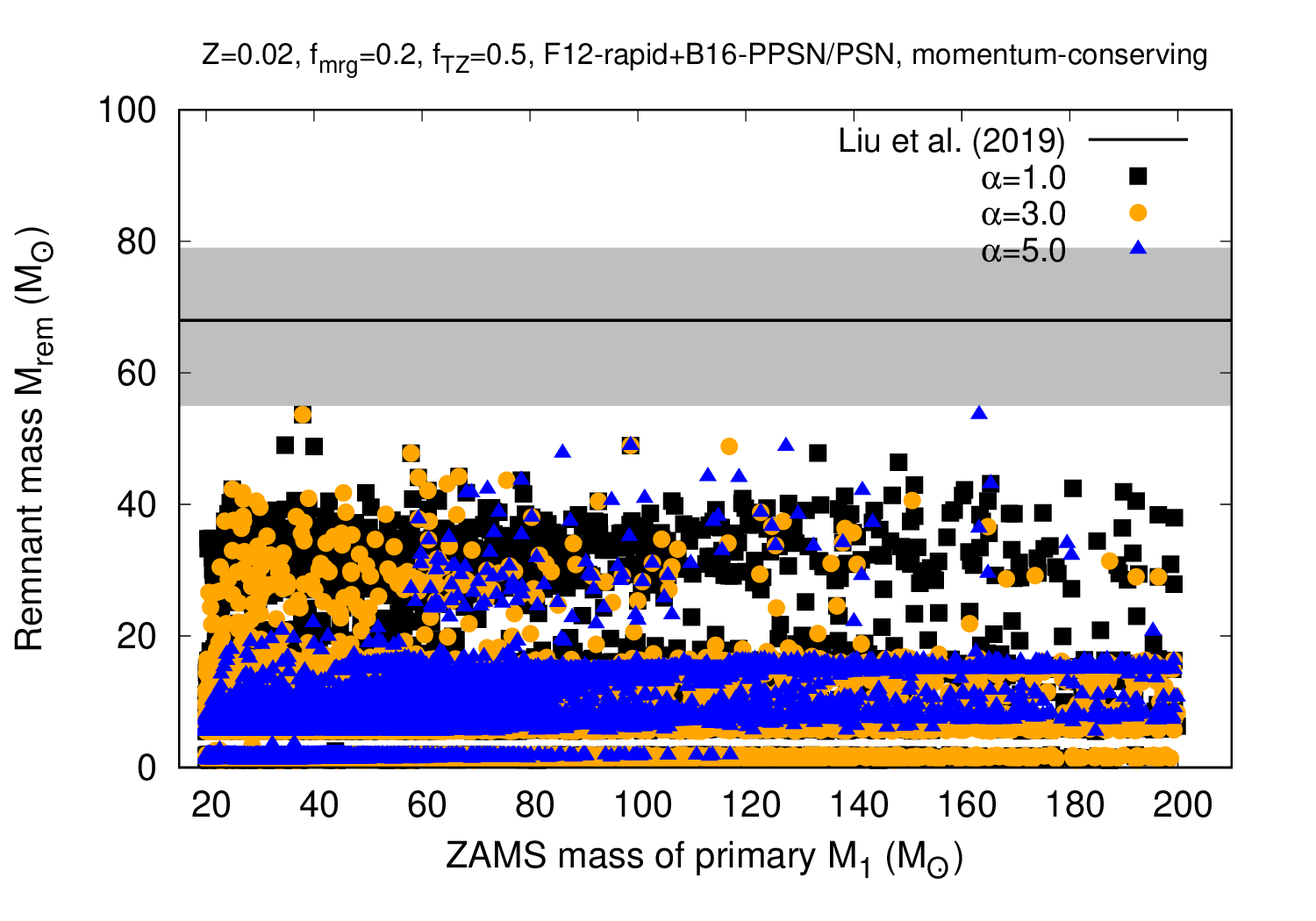}
\hspace{-0.5 cm}
\includegraphics[width=6.0cm,angle=0]{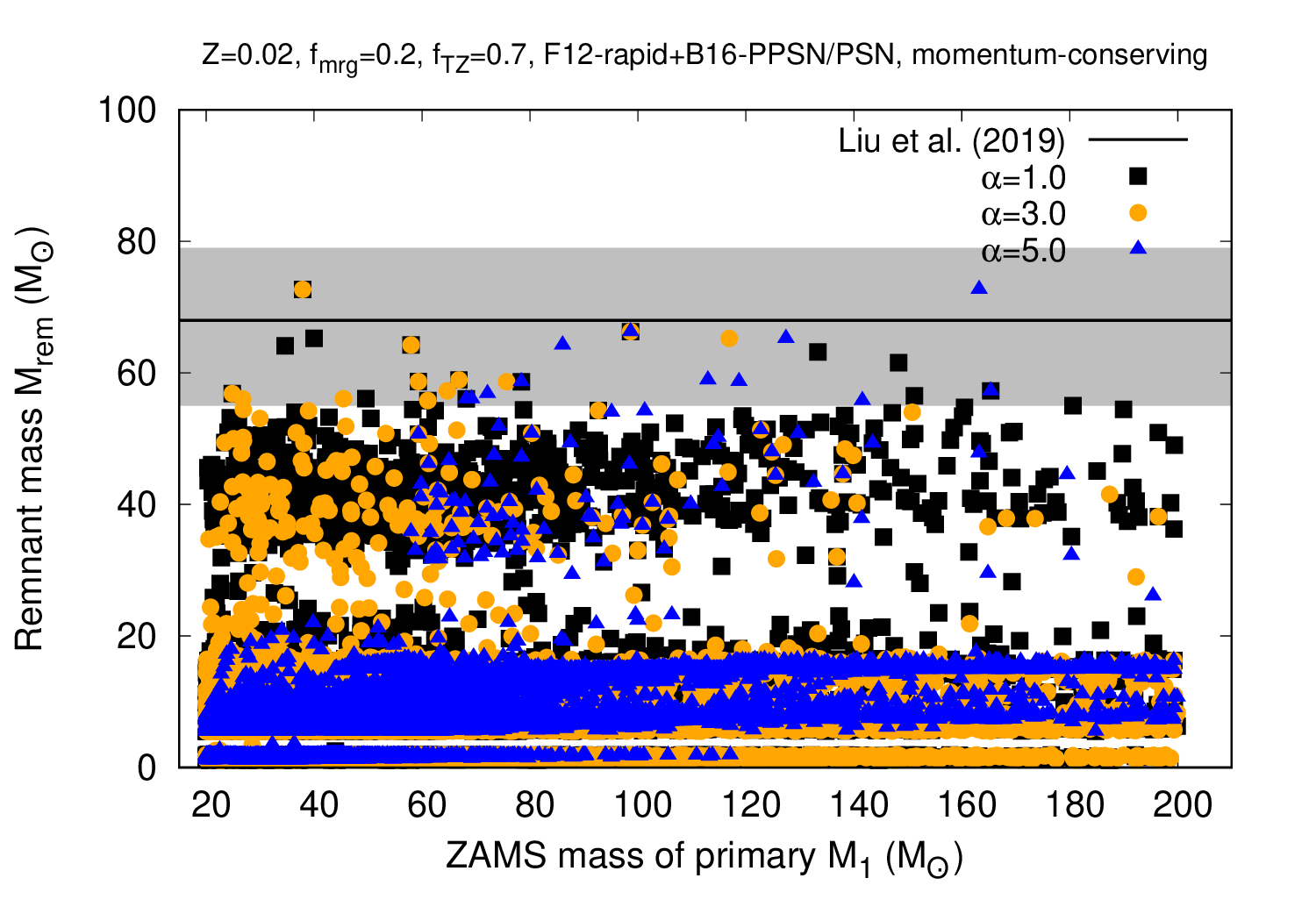}
\hspace{-0.5 cm}
\includegraphics[width=6.0cm,angle=0]{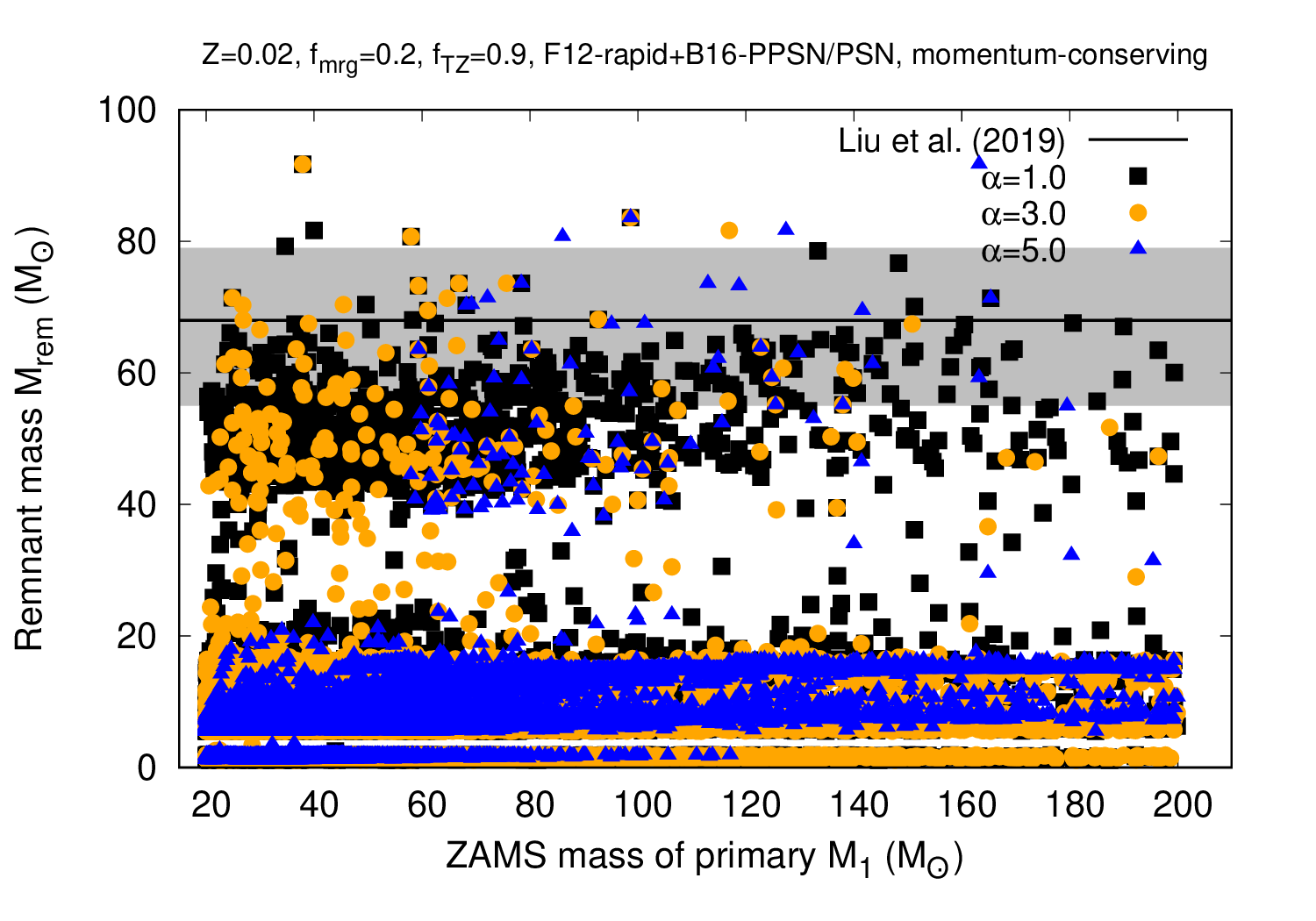}\\
\includegraphics[width=6.0cm,angle=0]{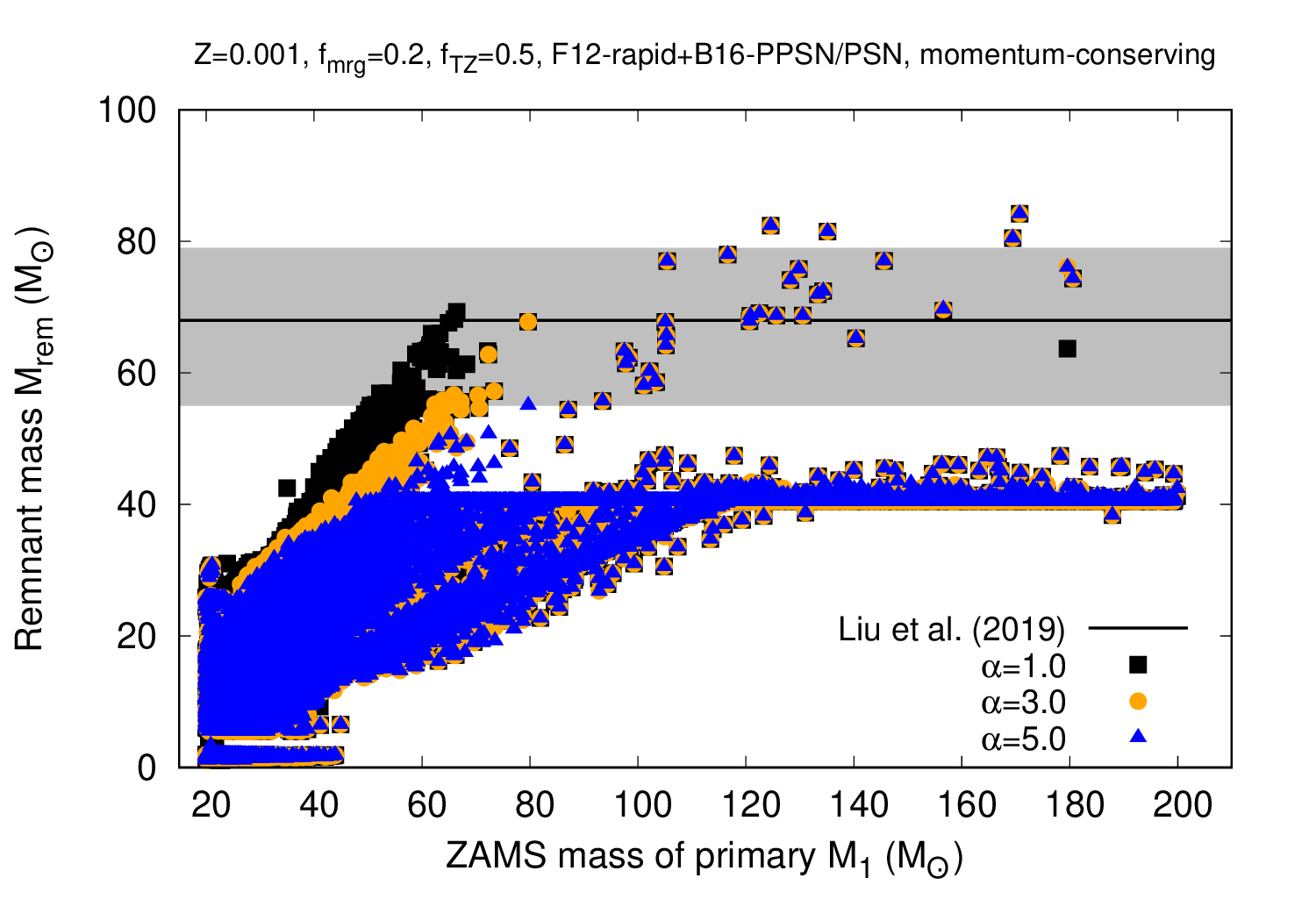}
\hspace{-0.5 cm}
\includegraphics[width=6.0cm,angle=0]{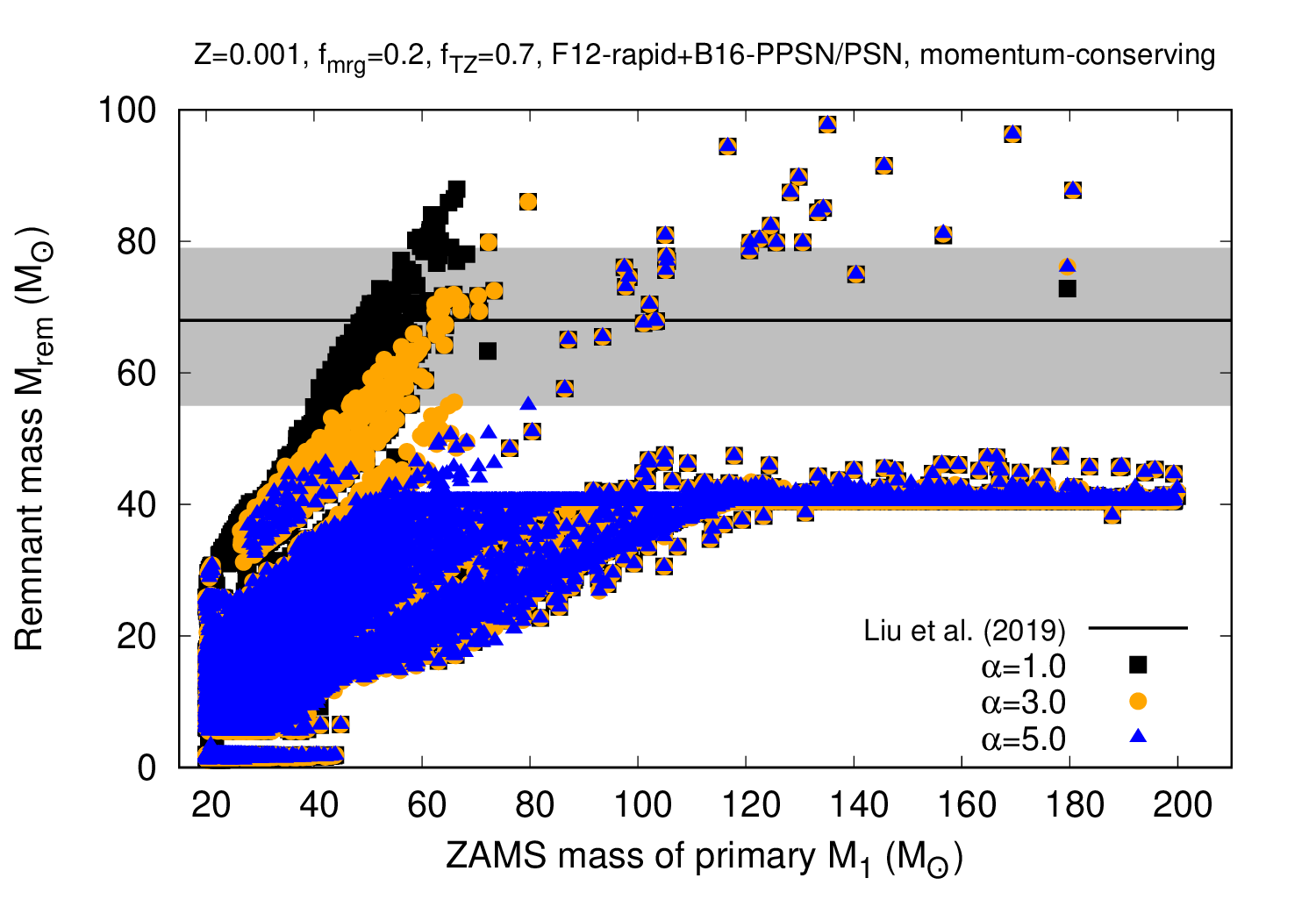}
\hspace{-0.5 cm}
\includegraphics[width=6.0cm,angle=0]{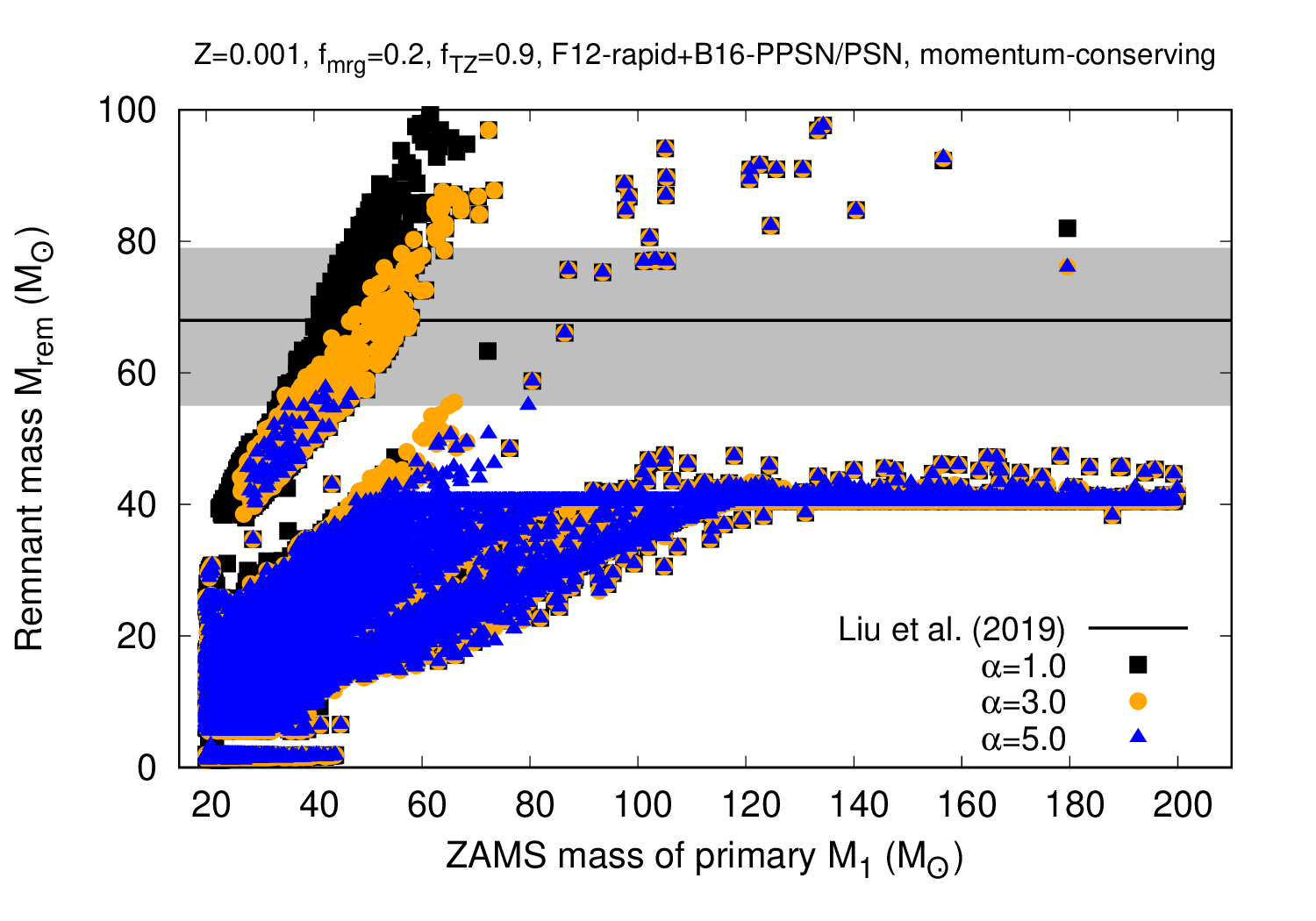}
	\caption{Zero age main sequence (ZAMS) mass-remnant mass relations for the massive-binary population evolution
	modelled in this work (comprising $10^4$ massive binaries; see Sec.~\ref{method}),
	for the different ``black hole Thorne-Zytkow object accretion fraction'', $\ftz$,
	star-star merger mass loss fraction, $\fmrg$,
	and common envelope (CE) efficiency parameter $\alpha$ as indicated in the panels' legends. For each panel,
	when a binary yields a single remnant, its primary's (the member with larger ZAMS mass)
	ZAMS mass is plotted along the x-axis. If a binary evolves to yield two remnants then
	the ZAMS masses of their parent stars are correspondingly plotted along the x-axis.
	The evolutions shown here assume the F12-rapid remnant formation model
	along with the PPSN/PSN model of B16 (Sec.~\ref{method}) except for those
	on the second row which correspond to the F12-delayed remnant model (Sec.~\ref{method}).
	All panels correspond to the metallicity $Z=\Zs=0.02$ except for those on the fourth row for which
	$Z=0.05\Zs$. The inferred black hole mass of
	LB-1 binary (black, solid line) and its uncertainty (90\%; grey, shaded region)
	by \citet{Liu_2019} are indicated on each panel. See Fig.~\ref{fig:fzero} for the
	ZAMS mass-remnant mass relation for the corresponding single stellar evolution.}
\label{fig:fpos}
\end{figure*}

\section{Method: population synthesis with the updated $\bse$}\label{method}

The binary-evolution program $\bse$ \citep{Hurley_2002} and its single-stellar evolutionary
counterpart $\sse$ \citep{Hurley_2000} are utilized to evolve the massive binary population
and the single massive stars (see below). $\sse$ and $\bse$ are fast, semi-analytical, recipe-based
programs that are widely utilized for stellar
population synthesis. They also serve as stellar evolution engines in widely used
N-body simulation programs such as $\nbsix$, $\nbseven$, $\nbpp$, {\tt MOCCA}, and {\tt CMC} 
\citep{2003gnbs.book.....A,2012MNRAS.422..841A,Wang_2015,Hypki_2013,Joshi_2000}.
$\sse$ and $\bse$ share the same single stellar evolution prescriptions
but $\bse$ incorporates additional recipes describing the binary-interaction physics
(tidal interaction, mass transfer, common-envelope evolution, coalescence, general-relativistic
orbit inspiral) and the binary-orbital mechanics; see \citet{Hurley_2002} for the details.
Here, an amended version of the original $\bse$, as described in \citet{Banerjee_2020},
is utilized.

To summarize, the amendments are in the $\sse$ sector with updates of the stellar wind
according to the recipes of \citet[hereafter B10]{Belczynski_2010}, of the stellar remnant
formation according to the ``rapid'' and ``delayed'' prescriptions of \citet[hereafter F12]{Fryer_2012},
and the implementations of pulsation pair-instability supernova (hereafter PPSN) and
pair-instability supernova (hereafter PSN) recipes as per \citet[hereafter B16]{Belczynski_2016a}.
Material fallback and neutrino mass loss (neutrino mass loss according to \citealt{Lattimer_1989} for an NS remnant,
assumed 10\% here for a BH remnant) are taken into account in implementing
the final NS or BH remnant mass. The fallback fraction is also
explicitly considered while implementing the natal kick of an NS or a BH based on the
``momentum-conserving'' principle as in \citet[hereafter B08]{Belczynski_2008}.
Alternatives of the momentum-conserving kick, \eg, ``collapse-asymmetry-driven'' and
``neutrino-emission-driven'' kicks are also kept as possibilities. The
binary-evolution physics remains the same as in \citet{Hurley_2002} along with its
subsequent amendments that are available in the public versions of $\bse$.
The original
$\sse$ recipes of \citet{Hurley_2000} and its earlier amendments are also retained and
the above, newest recipes can be opted for via appropriate option flags. That way, $\sse$ and
$\bse$ now offer a wide range of situations for exploring in stellar-evolutionary
population synthesis that are based on current understandings in stellar evolution and remnant formation,
being at par and in near-perfect agreement \citep{Banerjee_2020}
with contemporary population synthesis programs
such as {\tt StarTrack} \citep{Belczynski_2008}. Similar ingredients are
available also in the $\bse$-derivative {\tt MOBSE} \citep{Giacobbo_2018}
and the triple evolution program {\tt TrES} \citep{Toonen_2016}.

Two additional parameters are introduced in the binary population evolution conducted here.     
First, a constant mass-loss fraction, $\fmrg$, with respect to the less massive member
(as of just before the merger) during a star-star merger process. Secondly, a constant
mass-accretion fraction, $\ftz$, onto a BH when it coalesces with a star, forming a ``BH
Thorne-Zytkow object'' (hereafter BH-TZO). The corresponding accretion fraction on to an
NS Thorne-Zytkow object \citep[hereafter NS-TZO]{TZ_1975} is assumed to be always zero,
as defaulted in $\bse$\footnote{These parameters are starightforwardly
implemented in the subroutine {\tt MIX}.}.
Of course, the constancy of $\fmrg$ and $\ftz$ is an oversimplification  
which quantities, in reality, would depend on poorly understood or explored details
of these processes. Here, they simply serve as convenient parametrizations of star-star merger mass
loss and BH-TZO accretion. Unless stated otherwise, the B10 stellar wind, F12-rapid remnant formation
plus B16-PPSN/PSN, and momentum-conserving natal kick recipes are applied throughout this work
(the one-dimensional natal kick dispersion of a $1.4\Ms$ NS is taken to be $265\kmps$; \citealt{Hobbs_2005}).
Based on hydrodynamic studies of stellar mergers (\eg, \citealt{Gaburov_2008,deMink_2013,DeMink_2014}),
$\fmrg=0.2$ (\ie, $\leq10$\% loss from the total mass budget during a star-star merger)
is adopted throughout. The cases of $\ftz=0.0$, 0.5, 0.7, and 0.9 are explored
and as well those of the common envelope (CE) efficiency parameter $\alpha=1.0$, 3.0 and 5.0.
The solar metallicity is taken to be $Z=\Zs=0.02$, as defaulted in $\sse$.
Also, as noted in \citet{Banerjee_2020}, the $\sse$ time step parameters of
$(\ptsone,\ptstwo,\ptsthree)=(0.001,0.01,0.02)$ are applied to achieve stability
and convergence.

With the above settings, $10^4$ massive, O-type binaries are evolved. The binaries
initially follow the orbital-period, eccentricity, and mass-ratio distributions of 
\citet{Sana_2011} that represent the observed population of O-star binaries in
young clusters and the field. The ZAMS masses of the
individual binary members are drawn from the \citet{Kroupa_2001} initial mass function (IMF)
over the mass ranges $20.0\Ms - 150.0\Ms$ or $20.0\Ms - 200.0\Ms$. Motivated by
the $>150\Ms$ initial mass inferred for single stars in young massive clusters
(\eg, \citealt{Crowther_2010}), the somewhat higher stellar upper mass limit, $\mmax$, is
considered along with its canonical value of $\mmax\approx 150\Ms$ \citep{Weidner_2004}.
The binary populations are generated with the {\tt McLuster} program \citep{Kuepper_2011}.

\section{Results}\label{res}

Fig.~\ref{fig:fpos} shows the resulting ZAMS mass - remnant mass relations for the different
values of $\ftz$ and $\alpha$ as indicated on the panels. On each panel,
when a binary yields a single remnant, its primary's (the member with larger ZAMS mass)
ZAMS mass, $M_1$, is plotted along the x-axis. If a binary evolves to yield two remnants then
the ZAMS masses of their parent stars are correspondingly plotted along the x-axis.
The BH mass and its uncertainty (90\%), as inferred in the observations of LB-1 by \citet{Liu_2019},
are also shown on each panel. It can be seen that for $\ftz=0.7$, BHs up to
$\approx70\Ms$ are formed at $Z=0.02$ for both the $\mmax=150\Ms$ and $200\Ms$ binary populations, for
both F12-rapid and F12-delayed remnant-formation models, and for
$\alpha=1.0$, 3.0, and 5.0. For $\ftz=0.9$ and $Z=0.02$,
the maximum BH mass exceeds $80\Ms$ ($90\Ms$) for $\mmax=150\Ms$ ($\mmax=200\Ms$).
On the other hand, when $\ftz=0.5$ and $Z=0.02$ the BH masses do not reach even the lower bound of the LB-1
value ($55\Ms$) for $\mmax=150\Ms$ and marginally reaches the value for $\mmax=200\Ms$. 
Overall, with a sufficiently large BH-TZO accretion ($\gtrsim70$\%) the formation of LB-1-like BHs can
happen even at solar-like metallicities and with $\gtrsim90$\% accretion both the lower and
upper mass bounds of LB-1's BH can be easily accommodated.

At low metallicities,
BH masses can grow even larger. For example, at $Z=0.001$ (typical metallicity of Milky Way GCs),
the LB-1 BH mass can be covered even at $\ftz=0.5$ and for $\ftz=0.7$ and 0.9, BH masses exceed $100\Ms$
(see Fig.~\ref{fig:fpos}, fourth row).
Note that in all these evolutionary calculations,
the B10 wind is applied at its full unlike in recent works such as that of \citet{Belczynski_2020b}
where a reduced wind is applied (but see below). These exotic, high BH masses
at solar and lower metallicities occur due to the assumed significant BH-TZO accretion.
This is clear from Fig.~\ref{fig:fpos} as one scans the panels from left to right (increasing $\ftz$)
along a particular row. With such $\ftz$, although the most massive BHs form via BH-TZO accretion,
BBH and NS-BH mergers also contribute to the overall, broad mass spectrum of the BHs that is
derived from the binary-population evolution (see below). 

It would be worthwhile to investigate what sort of binary interaction(s) lead to such exotic
BH masses, especially at the solar metallicity. Examples 1a and 1b of Appendix~\ref{example}
demonstrate (via standard $\bse$ summary output) such evolutionary channels whose final outcomes are a single BH of
$68.8\Ms$ and $91.7\Ms$ respectively. Here, the more ZAMS-massive member (the primary) starts
a Case-A mass transfer as it expands near the end of its MS lifetime, it being
inside a close binary. The mass
transfer causes the secondary to gain mass significantly, thereby rejuvenating its
hydrogen fuel (complete mixing is assumed in $\bse$; see \citealt{Hurley_2002})
and prolonging its MS life (the secondary essentially
becomes a blue straggler). This continues until the mass donor has lost
all of its hydrogen envelope (due to wind loss plus the mass transfer) to become
a helium main sequence star ({\tt K2}$=7$; a Wolf-Rayet star) and, therefore, shrinks to a large extent,
detaching the binary.
Note that although the mass ratio has reversed by now,
we will continue to refer to the more ZAMS-massive member as the primary. Finally,
when the primary's He-core becomes a low-mass BH, it receives a moderate natal kick
whose value (as well as the BH's mass at birth of $5.7\Ms$) is determined by the (partial; $\approx60$\%
by mass)
fallback (see the kick information at the
beginning of Examples 1a \& 1b; F12-rapid remnant model and momentum-conserving kick
are assumed here).
This, in turn, causes the binary to become highly eccentric
(but still remain bound), leading to an in-orbit coalescence of the BH with its mass-gainer companion,
the formation of a BH-TZO, and, finally, the mass gain of the BH depending on $\ftz$.
In other words, the primary ``secures'' some of its mass with its secondary
by transferring material on to and rejuvenating the latter and then gains
it back, after becoming a low-mass BH, by merging with it.
In these examples, the
large mass gain of the BH is due to the adopted $\ftz=0.9$, as clarified at the end of
the corresponding $\bse$ outputs. Note that the moderate natal kick of the BH, due to the
partial fallback, plays an important role in this scenario by inducing a prompt merger;
a full NS-like kick would have likely disrupted such a binary while with too small
a kick (\eg, in a collapse-asymmetry-driven kick scenario; \citealt{Banerjee_2020}),
one would need to wait for the secondary to become a giant until it fills its Roche
lobe and the outcome would become $\alpha$-dependent ($\alpha=3.0$ is mentioned
for completeness but it has no real role in this evolutionary channel)
and, also, the stellar companion may lose a good part of its mass in the mean time
to its strong winds at the solar metallicity.

\begin{figure*}
\includegraphics[width=6.0cm,angle=0]{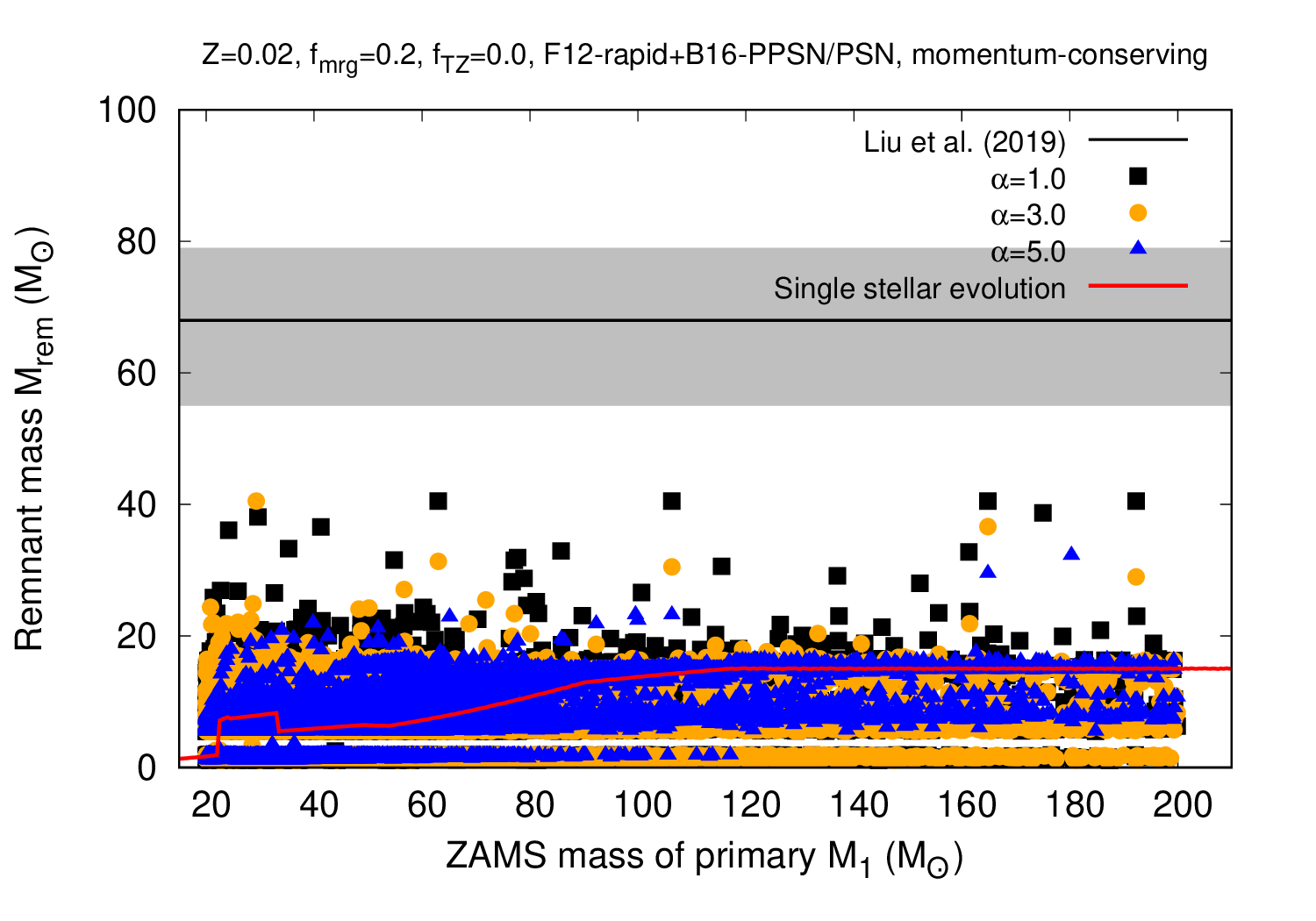}
\hspace{-0.5 cm}
\includegraphics[width=6.0cm,angle=0]{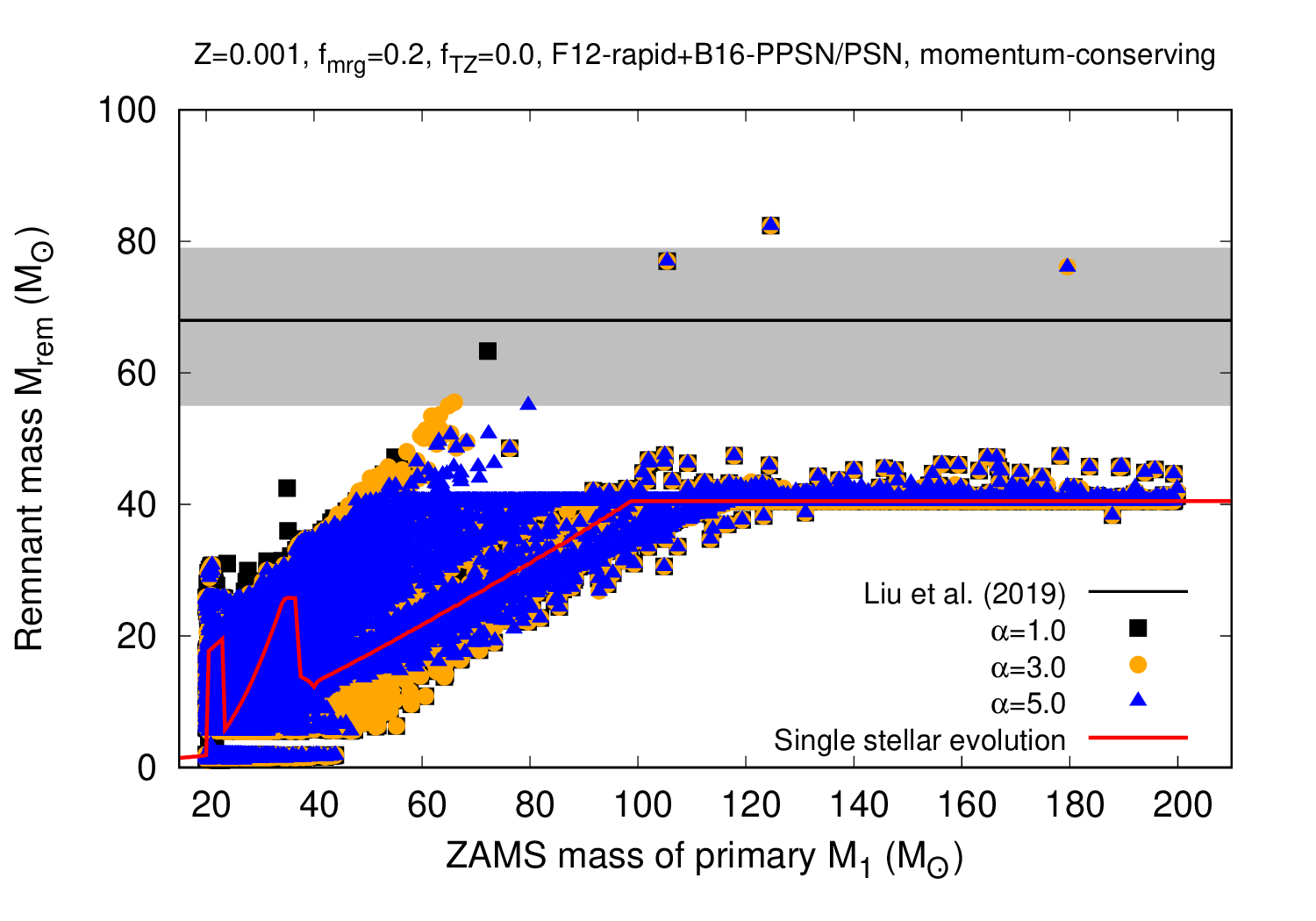}
\hspace{-0.5 cm}
\includegraphics[width=6.0cm,angle=0]{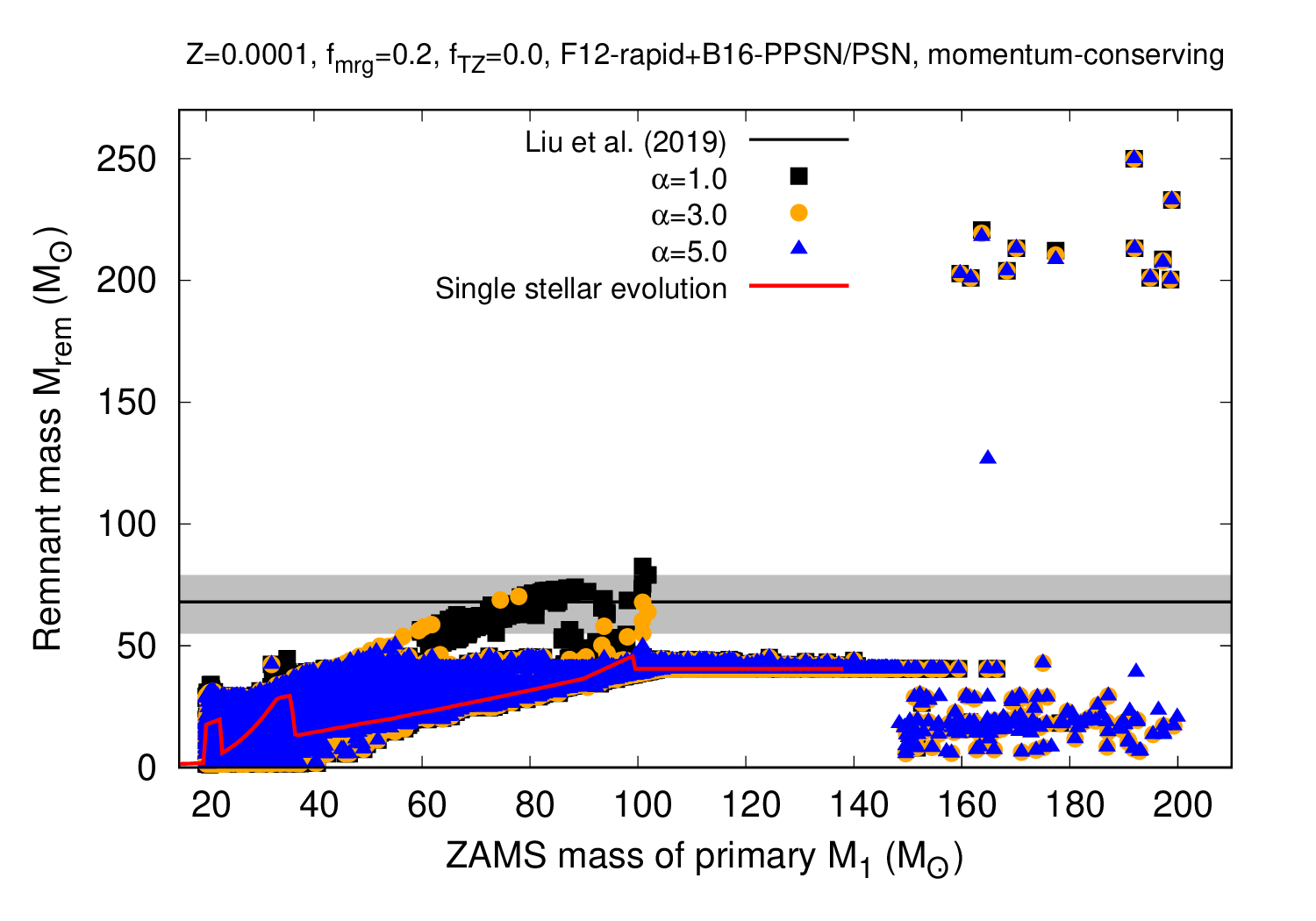}
	\caption{The same as in Fig.~\ref{fig:fpos} but with $\ftz=0.0$ and for $Z=\Zs=0.02$ (left panel),
	$Z=0.05\Zs$ (middle), and $Z=0.005\Zs$ (right). The red, solid line on each panel
	gives the ZAMS mass-remnant mass relation for the corresponding single stellar evolution.}
\label{fig:fzero}
\end{figure*}

Fig.~\ref{fig:fzero} shows the remnant masses without any BH-TZO accretion ($\ftz=0$) at $Z=0.02$,
0.001, and 0.0001, for the $\mmax=200\Ms$ binary population.
Here, LB-1-like and more massive (up to $\approx80\Ms$) BHs appear only at the
lower metallicities and in much fewer numbers. In the absence of
any BH-TZO accretion, these are the few, most massive BBH merger events. 
For $Z=0.02$, the BHs still
reach up to $\approx40\Ms$, these being derived, typically, from late-time merger
products due to Case-B mass transfer resulting in direct collapse or PPSN.
With $\ftz=0$, star-star mergers or mass accretion episodes from stellar companion
produce the majority of the over-massive BHs (with respect to the corresponding
single-star ZAMS mass-remnant mass relations as given by the red, solid line in the
panels of Fig.~\ref{fig:fzero}). The formation
of such over-massive BHs via stellar mergers at solar and lower metallicities have also been
inferred in other, recent studies, \eg, \citet{DeDonder_2004,Banerjee_2020,Spera_2019}.
The under-massive BHs, are, on the other hand, the end products of the net
mass givers in a binary evolution which are often the primaries.

A striking feature in the $\ftz=0$, $Z=0.0001$ case (Fig.\ref{fig:fzero}, right panel)
is the relatively low-mass ($5.0\lesssim\mrem\lesssim30.0$) BHs for $M_1\gtrsim150\Ms$,
mostly when $\alpha=3.0$ and 5.0, and as well the very high
mass ($\gtrsim200\Ms$) BHs over the same $M_1$ range when $\alpha=1.0$, 3.0, and 5.0.
The lower-mass BHs occur in wide, low-mass-ratio systems as in
Example 2a of Appendix~\ref{example}. Here, the onset of mass transfer across the components of widely
different masses leads to a CE phase and, subsequently, the ejection
of the H-envelope of the donor primary, owing to the efficient
energy deposition on to the envelope with $\alpha=3.0$. This results in a tight, detached, symbiotic
binary between a helium main sequence (or a naked-helium) primary and the main-sequence secondary. 
The He-star member, being in between the He-core mass range for PSN ($65.0\Ms-135.0\Ms$;
see \citealt{Belczynski_2016a,Woosley_2017}), leaves no remnant (its wind mass loss is small
due to the low $Z$) while the
secondary evolves into a $19\Ms$ BH. When $\alpha=1.0$ (Example 2b of Appendix~\ref{example}),
the CE ejection fails (less energy deposition on to the envelope), leading to a merger
and the merged star undergoes PSN so that no remnant is left at all. With
$\alpha=3.0$ and the higher $Z=0.001$ (Example 2c), the evolutionary path is similar to   
that of Example 2a but, owing to higher wind mass loss, the primary
undergoes a PPSN and the secondary becomes an NS instead. Before the NS formation,
a stable mass-transfer phase
occurs from the secondary to the BH primary, increasing the latter's mass slightly.  
The binary finally disrupts due to the high natal kick of the NS. 
It is this type of mass gain via Roche lobe overflow (as opposed to BH-TZO accretion) 
that leads to the small, preferentially upward spread around the PPSN plateau
in the initial-final plots for the low metallicities
(Fig.~\ref{fig:fpos}, fourth row; Fig.~\ref{fig:fzero}, middle and right panels).

On the other hand, the very massive, $\gtrsim200\Ms$ BHs are the outcomes of
tight, very massive binaries leading to direct collapse
above the PSN mass gap \citep{Spera_2015,Belczynski_2016a,Woosley_2017,Giacobbo_2018}
as in Example 3a of Appendix~\ref{example}.
The star-star merger product, in this case, is sufficiently
massive and its wind mass loss is low enough (due to low $Z$) to yield a BH above the PSN gap
(or the ``upper'' mass gap; \cf Fig.~2, bottom panel of
\citealt{Banerjee_2020}). 

\begin{figure*}
\includegraphics[width=8.0cm,angle=0]{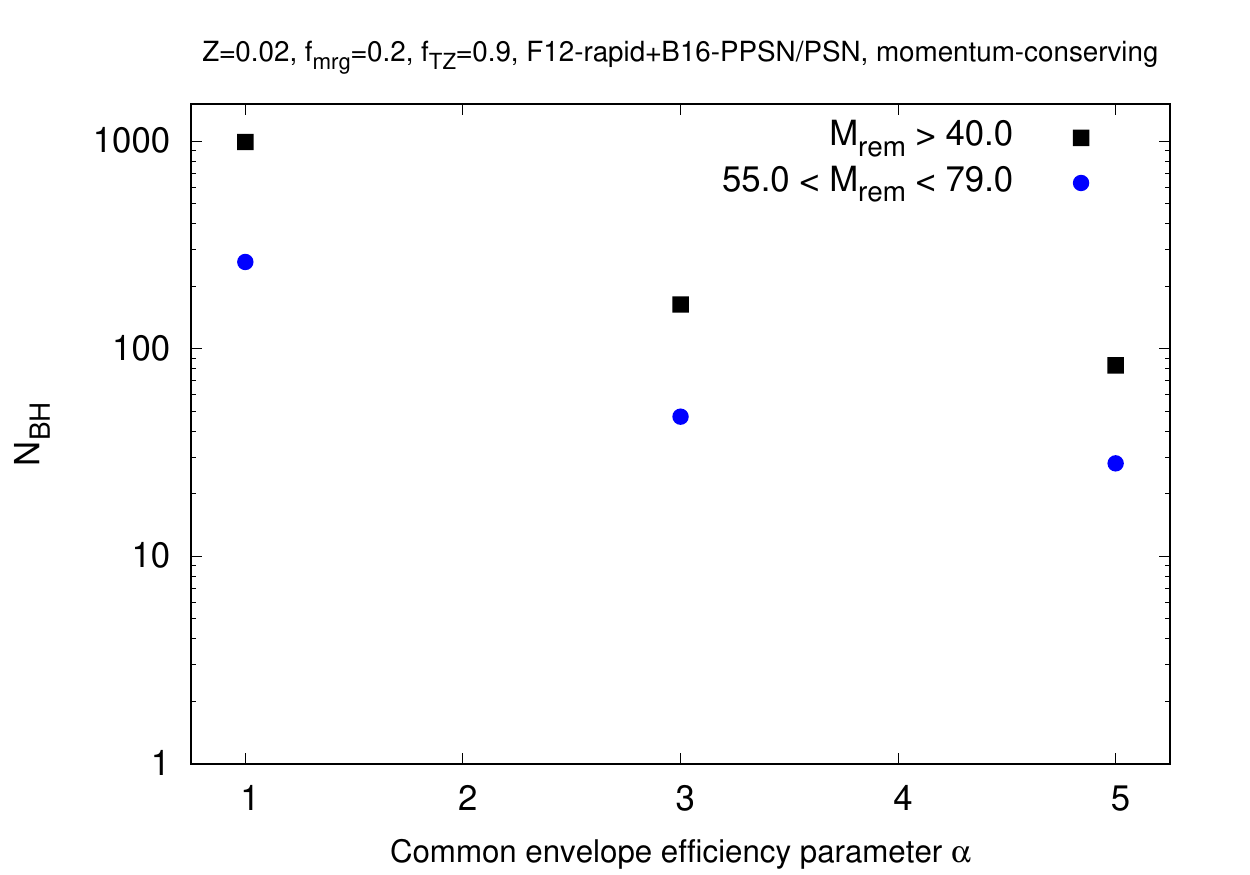}
\includegraphics[width=8.0cm,angle=0]{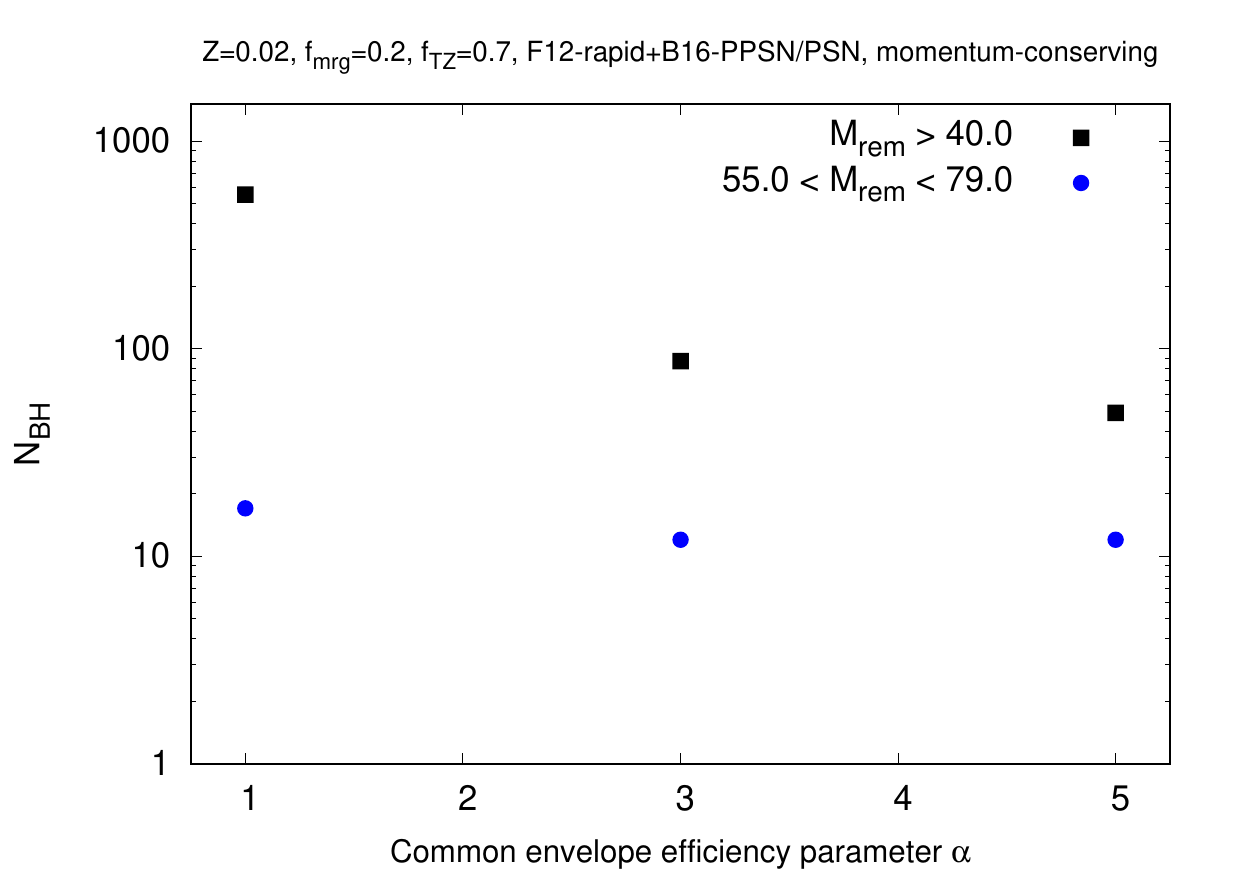}\\
\includegraphics[width=8.0cm,angle=0]{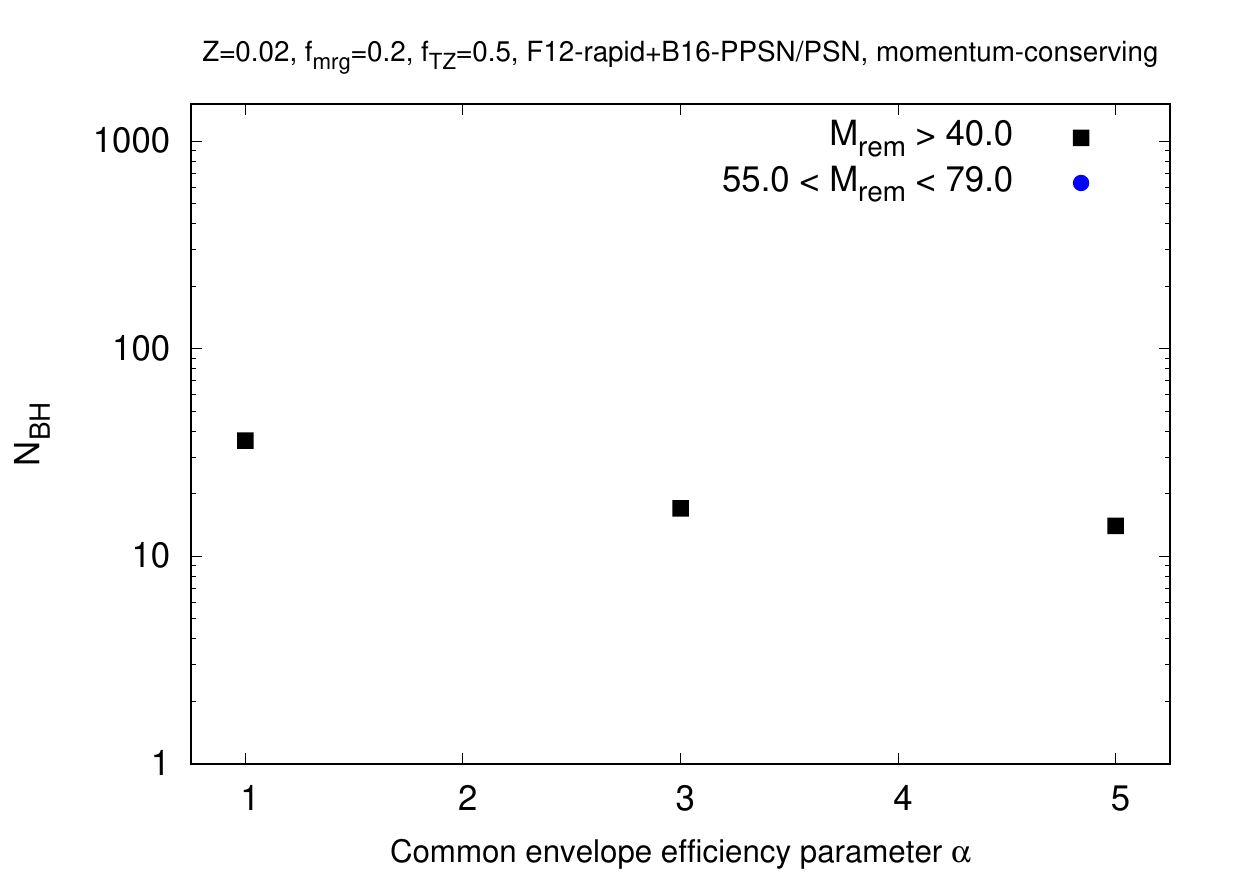}
\includegraphics[width=8.0cm,angle=0]{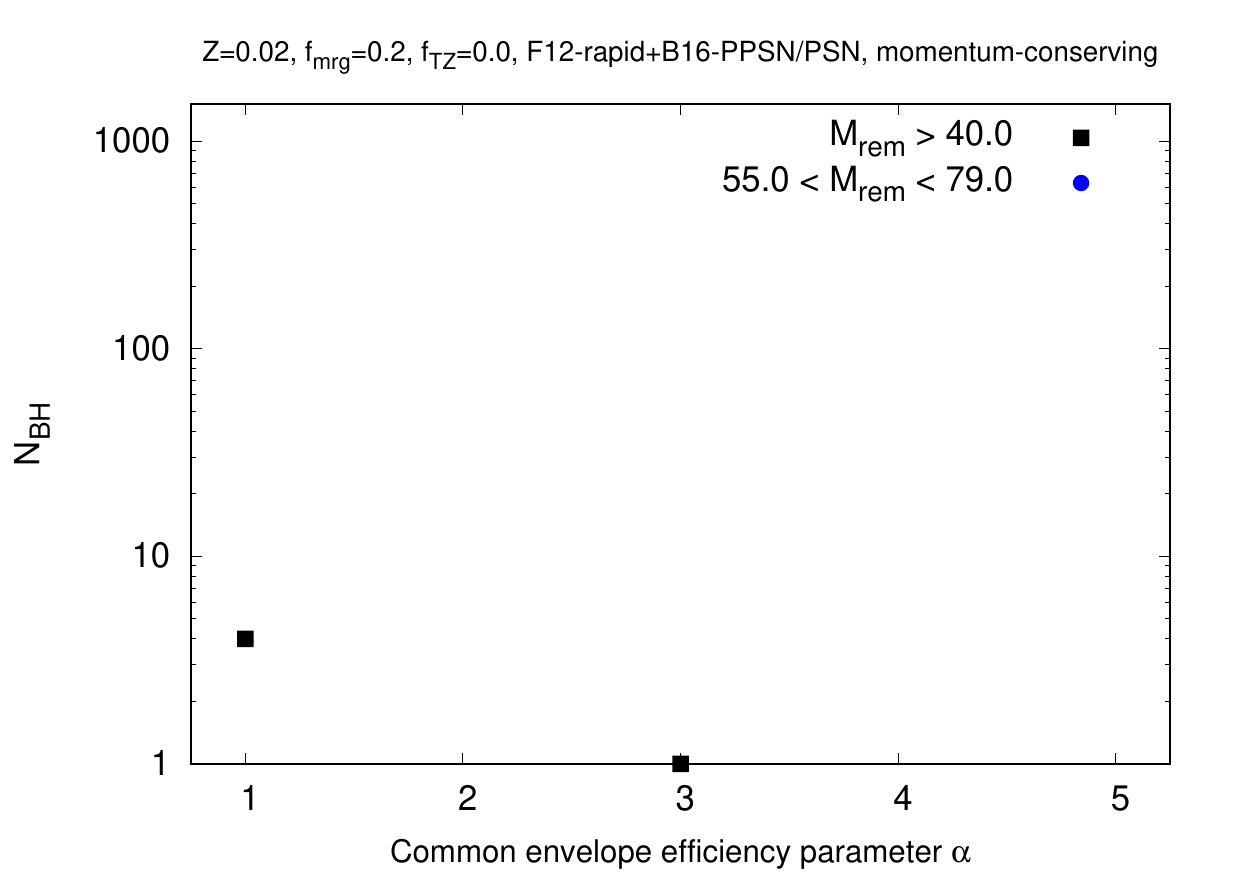}
	\caption{The total numbers of BHs, $\nbh$, formed, with $\mrem>40.0\Ms$ (the generally-accepted
	mass of a PPSN remnant BH; black, filled squares) and with $55.0<\mrem<79.0$
	(the mass range of the BH inferred in LB-1; blue, filled circles), out of
	the $10^4$ massive binaries (Sec.~\ref{method}) as a function of the CE efficiency parameter $\alpha$.
	Here, the cases with
	$\ftz=0.9$ (top-left panel), 0.7 (top-right), 0.5 (bottom-left), and 0.0 (bottom-right), for
	$Z=\Zs=0.02$ and $\fmrg=0.2$, are shown. These panels correspond to the outcomes
	from the population of binaries with components having ZAMS mass up to $200\Ms$ (Sec.~\ref{method}).}
\label{fig:bhnum}
\end{figure*}

Fig.~\ref{fig:bhnum} shows the number of BHs, at $Z=0.02$, with $\mrem>40.0\Ms$, the generally-accepted
lower mass limit of a PPSN remnant BH \citep{Belczynski_2016a,Woosley_2017},
and with $55.0<\mrem<79.0$, the mass range of the BH in LB-1 as inferred by \citet{Liu_2019},
as functions of $\ftz$ (the different panels) and $\alpha$ (the x-axes). As expected from
Figs.~\ref{fig:fpos} \& \ref{fig:fzero} and the discussions
above, $\ftz\gtrsim0.7$ is necessary to produce such
massive BHs. The moderately negative $\alpha$-dependence is due to the fact that a lower $\alpha$
generally aids in star-star and star-remnant mergers due to the correspondingly longer H-envelope lifetime during a
CE phase.
Out of the $2\times10^4$ stellar members ($10^4$ binaries),
the fractional relation with the number of
$55.0<\mrem<79.0{\rm~}(\mrem>40.0\Ms)$ BHs formed is $\approx1.5\times10^{-2}{\rm~}(5.0\times10^{-2})$,
for $\ftz=0.9$ and $\alpha=1.0$. For $\ftz=0.7$, $\alpha=1.0$ it
is $\approx1.0\times10^{-3}{\rm~}(2.5\times10^{-2})$. These fractions represent lower limits
since only $10^4$ binaries are evolved in this work; a more robust estimate, which is reserved for
a future study, would require $\sim10^7$ binaries. Note that these numbers
correspond to the stellar mass range of $20\Ms-200\Ms$ (for the $\mmax=150\Ms$ binary population,
the numbers are marginally lower) and assuming 100\% binary fraction over this mass range. With
respect to the full, standard IMF, these fractions would easily be $\sim10^{-2}$ factors lower and even
lower for $<100$\% binary fractions among O-type stars (the present-day, observed O-star
binary fraction is $\approx50-70$\%; \citealt{Sana_2011,Sana_2013}).

\begin{figure*}
\includegraphics[width=6.0cm,angle=0]{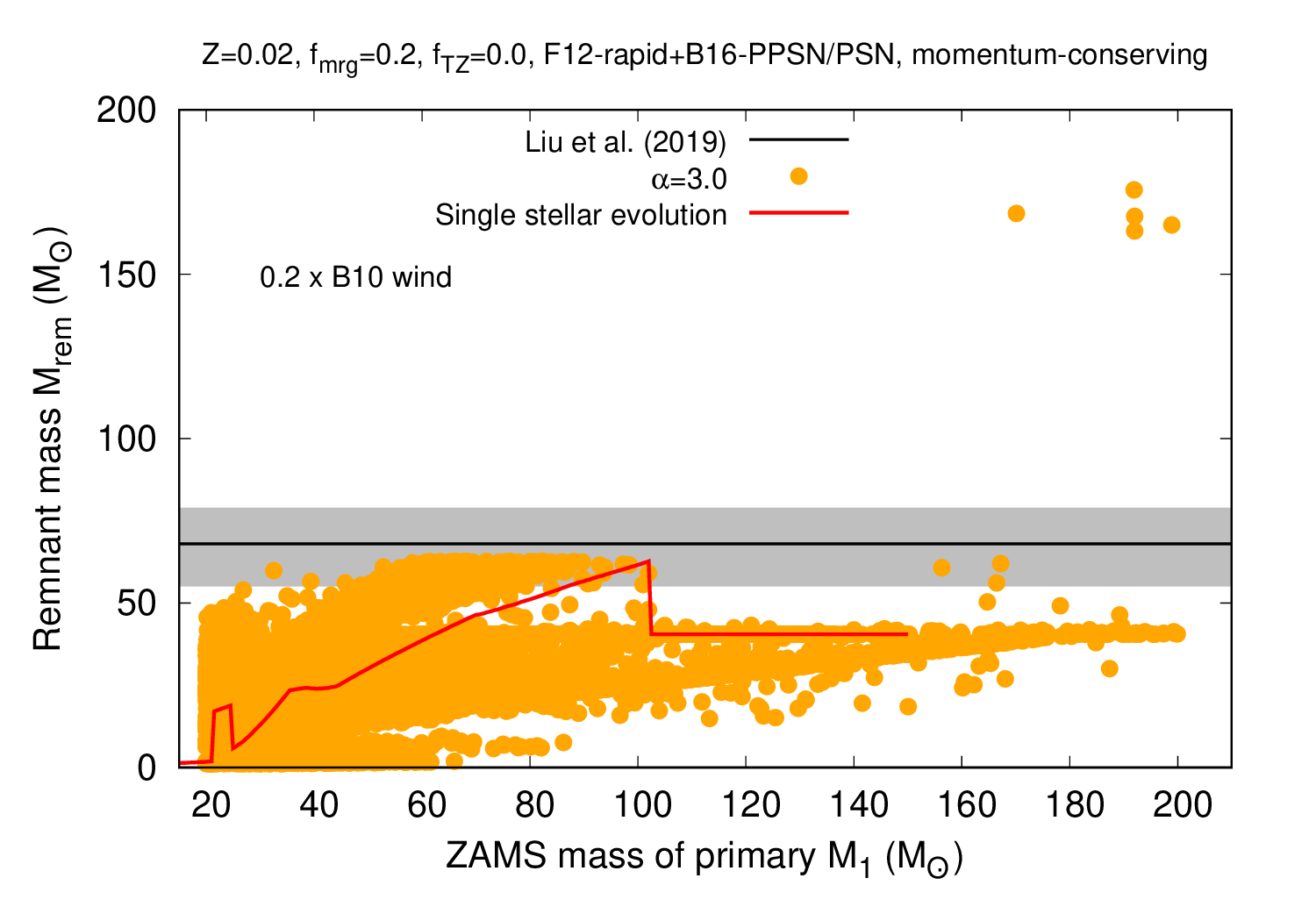}
\hspace{-0.5 cm}
\includegraphics[width=6.0cm,angle=0]{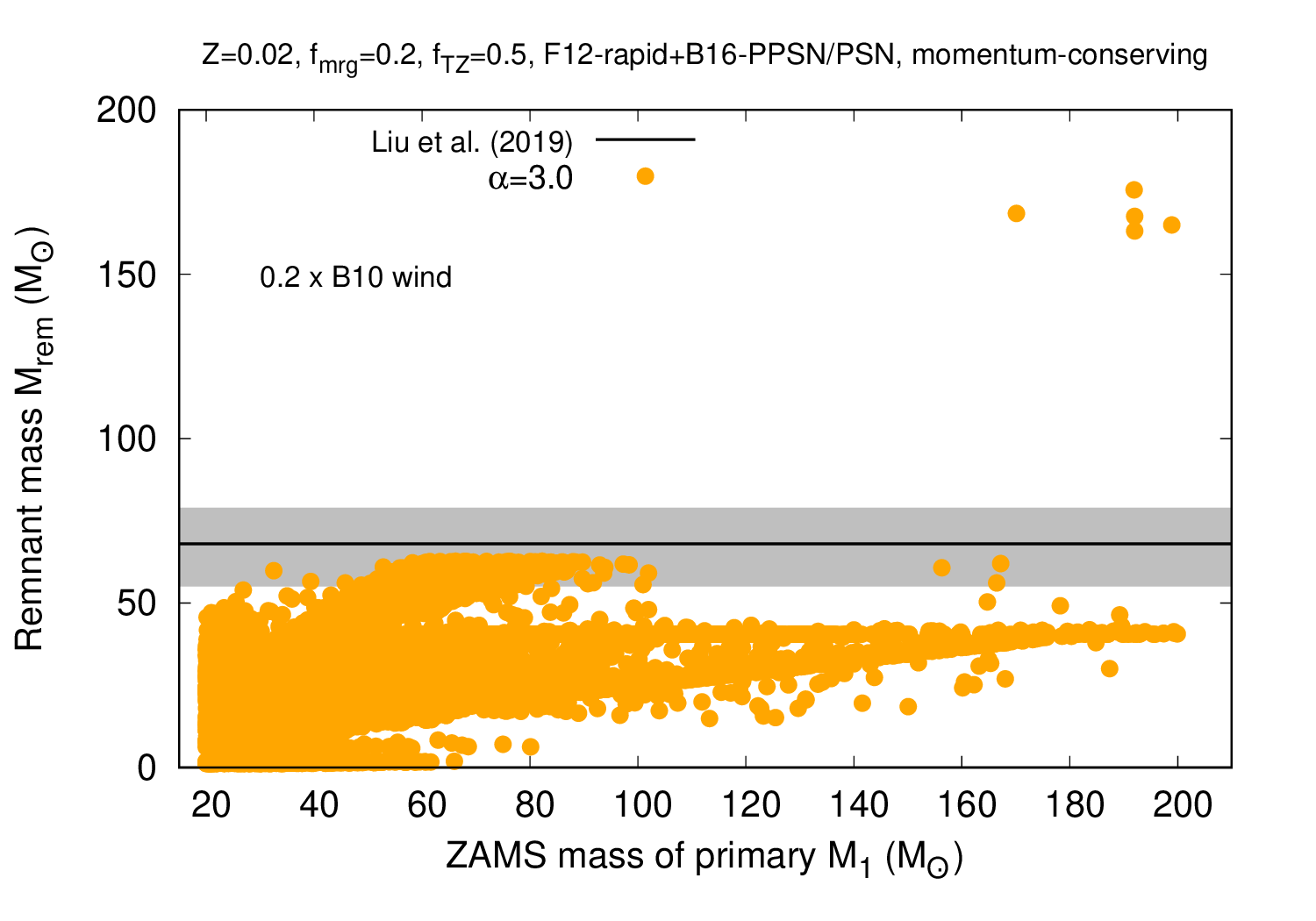}
\hspace{-0.5 cm}
\includegraphics[width=6.0cm,angle=0]{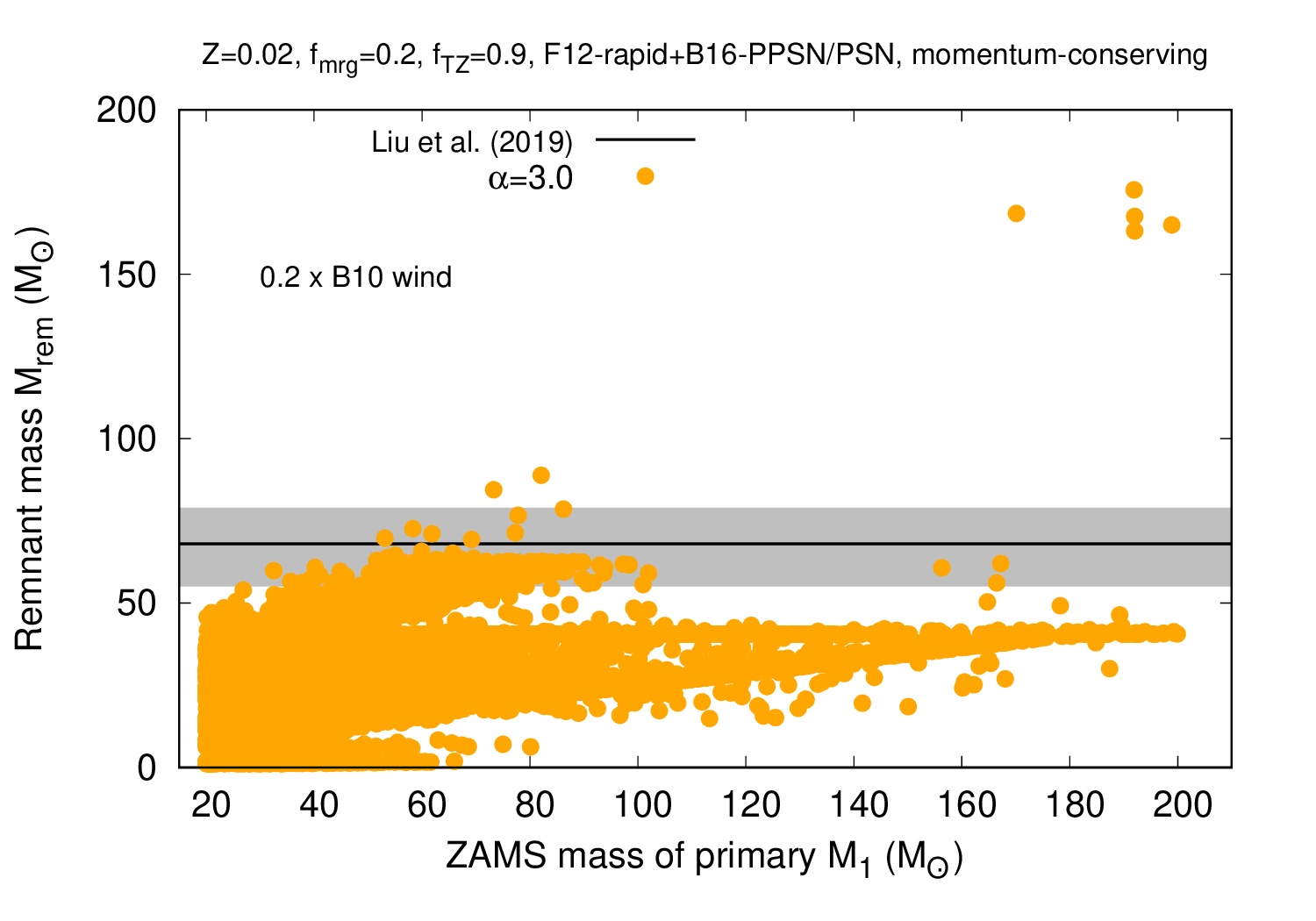}
\caption{The same as in Fig.~\ref{fig:fpos} but with the stellar wind reduced to 20\% of the standard B10
	value (Sec.~\ref{method}) as in \citet{Belczynski_2020b}. Here, the cases with
	$\ftz=0.0$ (left panel), 0.5 (middle), 0.9 (right), for $\alpha=3.0$,
	$Z=\Zs=0.02$, and $\fmrg=0.2$, are shown. The left panel includes the ZAMS mass-remnant mass relation
	for the corresponding single stellar evolution (the red, solid line).}
\label{fig:wcon02}
\end{figure*}

Finally, Fig.~\ref{fig:wcon02} shows the remnant outcomes, for $Z=0.02$ and $\alpha=3.0$, when the B10 wind in
the evolutionary models is reduced by an arbitrary factor of 0.2. This corresponds
to the situation explored recently by \citet{Belczynski_2020b} with
single stellar evolution using {\tt StarTrack} \citep{Belczynski_2008,Belczynski_2016a}
and {\tt MESA} \citep{Paxton_2011,Paxton_2015} programs. The corresponding single-star initial-final
relation is shown on the left panel of Fig.~\ref{fig:wcon02} (the red, solid line) which resembles
that from \citet{Belczynski_2020b} (their Fig.~2; the maximum BH mass is slightly higher
in their case since they assume 1\% neutrino mass loss during BH formation as opposed
to that of 10\% here). As can then be expected, the LB-1 BH mass can be reached  
even with $\ftz=0.0$. The very massive, $>150\Ms$ BHs in the panels of Fig.~\ref{fig:wcon02} 
are above-PSN-gap BHs as in Fig.~\ref{fig:fzero} (right panel) that is discussed above.
They are produced through early mergers of tight, massive binaries where the poor mass loss of
the merger product results in a BH above the PSN gap, as demonstrated in Example 3b
of Appendix~\ref{example}. If one resets to the full B10 wind, then the same merger
product leads to just a $\approx15\Ms$ BH (Example 3c of Appendix~\ref{example}) as one
expects from such massive ZAMS stars at the solar metallicity (see, \eg, Fig.~1, bottom panels
of \citealt{Banerjee_2020}). Note that since the merger happened right at
the beginning due to the overlapping initial separation, the $\approx296\Ms$ merger product, in this example,
is essentially a ZAMS star.

\section{Discussions and outlook}\label{discuss}

The formation of a merger product (a BH-TZO) between a few-$\Ms$ BH and a 10s-of-$\Ms$
stellar object at solar metallicity as in Examples 1a and 1b (Appendix~\ref{example}), staring
from a close, massive binary, is, by itself,
not exotic and depends mainly on the essentials of binary evolution and 
orbital mechanics. The details of the $\bse$'s underlying stellar structure
do not play a key role up to the BH-TZO formation point. Also,
such an outcome happens with the full B10 wind. The exotic
aspect is the large amount of mass accretion fraction, $\ftz$,
that had to be imposed to grow the BH up to $50\Ms-80\Ms$, the mass limits of the
LB-1 BH by \citet{Liu_2019}. It is unclear if $\gtrsim70$\% of mass can be
accreted onto the BH from its gaseous cocoon and, even if so, how much time
the accretion would take. With $\lesssim50$\% BH-TZO accretion, $40\Ms-50\Ms$
BH would still form, but with a very low probability (Fig.~\ref{fig:bhnum}, Sec.~\ref{res}). 

However, looking from another angle, a large BH-TZO accretion fraction
is analogous to the model of BH formation with a large fallback fraction on to a $\sim\Ms$
proto-remnant \citep[and references therein]{Fryer_2012}, as adopted in essentially all
contemporary population-synthesis and hydrodynamic approaches
(\eg, in $\bse$, {\tt MOBSE}, {\tt StarTrack}, and {\tt MESA} programs) for
the final remnant formation from an evolved stellar entity.
In that respect, the adoption of a large $\ftz$ is as reasonable as
BH formation with near-complete fallback of matter onto
a several times less massive proto-remnant. The fact that LIGO-Virgo
has observed BHs up to $\approx50\Ms$ \citep{Abbott_GWTC1}
indicates that such fallback-dominated BH formation somehow works out
and hence a large $\ftz$ may as well.

Along the same line of argument, the final BH out of
a BH-star merger product (which one would normally expect
to be of high spin parameter) can, in fact, be of low or
practically zero spin parameter, as recently inferred for
BH formation from stellar collapse using {\tt MESA}
\citep{Belczynski_2020}.
In an evolved star, the angular momentum of
the innermost radiative core is carried away
outwards (and expelled from the system via the wind)
almost entirely,
due to the core's strong magnetic coupling with
the outer regions caused by the twisting
of magnetic field lines threaded into it \citep{Fuller_2019}. 
Given that the BH inside a TZO is also surrounded
by a massive cocoon (as opposed to a BH gaining mass
from a stellar companion through an accretion disk
in which case its spin-up is almost guaranteed), 
its spin fate may be similar to that of a stellar core.
In the case of a BH core, the magnetic field threading
and the resulting angular momentum extraction from it
would happen relativistically \citep{Blandford_1977,Thorne_1986}.
General relativistic (GR) magnetohydrodynamic (GRMHD) studies are
necessary to better understand the fate of such a
BH-TZO.

With such BH-TZO accretion, BH mass can reach $\approx100\Ms$ at
GC-like metallicities (Fig.~\ref{fig:fpos}, fourth row;
Sec.~\ref{res}). Without any BH-TZO accretion (Fig.~\ref{fig:fzero}),
BHs barely reach $\approx40\Ms$ at $Z=0.02$ and $\approx80\Ms$ for GC-like 
and lower $Z$, these being outcomes of late-time star-star mergers or
BBH mergers (Sec.~\ref{res}). These channels operate as well with
finite $\ftz$, for all $Z$. The same applies for mass growth of BHs via accretion from
stellar companion.

This work does not address the question how a $\approx70\Ms$
BH would acquire a B-type companion. Although the near circular orbit
points to an origin from field binary evolution, the key challenge
in such a scenario is to avoid merger due to the evolutionary expansion
of the BH progenitor. Any evolved BH-progenitor star with H-rich envelope
would be much bigger in size than the $\approx300\Rs$ ($\approx80$-day)
orbit of LB-1 (if the LB-1's BH turns out to be of much lower mass, then
the orbit would have to be tighter but, depending on the inferred BH mass,
one may be able to resort to CE and envelope ejection). It is possible
that the BH progenitor evolves chemically homogeneously \citep{DeMink_2016,Marchant_2016},
maintaining a compact size throughout its lifetime, but then
the companion is too far away to circularize any primordial
eccentricity through tidal interaction. Although there is
evidence of chemical homogeneity in massive BH progenitors in the metal-poor
SMC \citep{Ramachandran_2019}, it is unclear whether the same would happen
for a Milky Way-like enrichment. The binary is
also too wide to induce chemical homogeneity in the BH progenitor
rotationally \citep{DeMink_2009,Marchant_2016}.

Alternatively, the BH can easily be exchanged into
a star-star binary in a close encounter inside a low-mass (massive) open cluster
\citep{Banerjee_2018}, which system can then become a member of the field
after the cluster has dissolved (it is ejected from
the cluster due to this or subsequent dynamical interaction).
This scenario faces difficulty in addressing the near-circular orbit
of LB-1, since a dynamically formed and/or ejected binary
would have an eccentricity drawn from the thermal distribution \citep{1987degc.book.....S}.
However, with an appropriately-high eccentricity, the orphaned BH-star binary
can become symbiotic, circularizing and tightening itself.
These possibilities will be investigated in a forthcoming work.
See \citet{Belczynski_2020b} for further possibilities.

This preliminary work has focussed on the most massive BHs formed out of a
population of massive binaries. Irrespective of whether LB-1's BH mass
would require a revision after follow-up observations or not, such a study
is interesting and contextual by its own right. The high (50\%-70\%) binary fraction
among the O-stars in present-day young massive and open clusters suggests
similar stellar population in GCs' progenitor clusters. Hence,
the properties (BH masses, spins) of dynamically-assembled BBHs and their GR mergers
in open and globular clusters would be influenced by the
pairing properties of the BHs' progenitor stars.
In the near future, larger sets of binary population will be evolved
and a detailed study of the resulting remnants' properties (masses
and spins) will be made.

\section*{Acknowledgements}

SB acknowledges the support from the Deutsche Forschungsgemeinschaft (DFG; German Research Foundation)
through the individual research grant ``The dynamics of stellar-mass black holes in
dense stellar systems and their role in gravitational-wave generation'' (BA 4281/6-1; PI: S. Banerjee).
The author acknowledges the discussions with Chris Belczynski, Mirek Giersz (CAMK, Warsaw),
Rainer Spurzem, and Peter Berczik (NAOC, Beijing). The motivation for implementing
a BH accretion in a BH-star merger product came from an unrelated discussion with Mirek Giersz.
SB acknowledges the generous support and efficient system maintenance of the
computing team at the AIfA.


\bibliographystyle{mnras}
\bibliography{bibliography/biblio.bib}

\label{lastpage}

\newpage

\appendix

\section{Examples of typical binary-evolutionary histories leading to massive black holes at solar and
sub-solar metallicities}\label{example}

In the following, individual examples from the present $\bse$ binary-evolutionary models, leading to
exotic BH masses, are provided. In all these cases, the F12-rapid remnant-formation model (including
B16-PPSN/PSN) and $\fmrg=0.2$ (Sec.~\ref{method}) are applied. The stellar wind mass loss is always
according to the B10 prescription except in Example 3a where 20\% of this wind is applied. 
The $\sse$ time step parameters of $(\ptsone,\ptstwo,\ptsthree)=(0.001,0.01,0.02)$ are applied
in all cases.

\onecolumn

\subsection*{Example 1a ($Z=0.02$, $\ftz=0.9$, $\alpha=3.0$):}
\begin{minipage}[h]{\textwidth}
\begin{verbatim}
 NS/BH formation (mechanism/fallback control) MASS KS FBFAC FBTOT MCO VKICK KMECH:
 5.7393168333929676   14  0.62576939152049149        5.3770187037699637
 7.3172588535468650        197.14296511276697    1

      TIME      M1       M2   K1 K2        SEP    ECC  R1/ROL1 R2/ROL2  TYPE
     0.0000   44.451   66.882  1  1       78.729  0.19   0.368   0.401  INITIAL 
     3.2806   41.324   54.557  1  1       78.914  0.00   0.648   1.001  BEG RCHE
     3.8856   44.232   48.716  1  2       75.743  0.00   0.777   1.465  KW CHNGE
     3.8903   48.963   43.958  1  4       72.412  0.00   0.733  81.084  KW CHNGE
     3.8918   59.374   33.534  1  4       81.898  0.00   0.577  88.638  BEG BSS 
     3.8942   74.135   18.751  1  4      158.043  0.00   0.266   0.450  END RCHE
     3.8978   74.143   18.674  1  7      158.022  0.00   0.266   0.032  KW CHNGE
     4.4266   70.247    9.919  1  8      182.610  0.00   0.241   0.022  KW CHNGE
     4.4535   70.045    5.739  1 14       96.990  0.90   0.424   0.000  KW CHNGE
     4.4535   70.045    5.739  1 14       96.935  0.90   0.424   0.000  CONTACT 
     4.4535   68.780    9.919 14 15        0.000  0.00   0.000  -1.000  COELESCE
 13000.0000   68.780    0.000 14 15        0.000 -1.00   0.000  -1.000  MAX TIME
\end{verbatim}
Note: with $\ftz=0.9$, the final BH mass $=5.739\Ms+(70.045\Ms\times0.9)=68.780\Ms$.
\end{minipage}

\subsection*{Example 1b ($Z=0.02$, $\ftz=0.9$, $\alpha=3.0$):}
\begin{minipage}[h]{\textwidth}
\begin{verbatim}
 NS/BH formation (mechanism/fallback control) MASS KS FBFAC FBTOT MCO VKICK KMECH:
 6.0403669764522627   14  0.63587200393794840        5.7115188627247360
 7.6300772770290646        196.20808369097909    1

      TIME      M1       M2   K1 K2        SEP    ECC  R1/ROL1 R2/ROL2  TYPE
     0.0000   39.218  163.208  1  1       95.036  0.20   0.366   0.506  INITIAL 
     2.2629   39.065  105.439  1  1       87.277  0.00   0.493   1.001  BEG RCHE
     3.2821   70.793   51.518  1  2       62.362  0.00   0.744   2.116  KW CHNGE
     3.2865   76.465   45.798  1  4       64.706  0.00   0.710 105.578  KW CHNGE
     3.2902  101.985   20.230  1  4      159.736  0.00   0.268   0.914  END RCHE
     3.2956  101.949   20.171  1  7      159.790  0.00   0.268   0.036  KW CHNGE
     3.3427  101.387   18.630  1  7      162.759  0.00   0.263   0.035  BEG BSS 
     3.8053   95.552   10.324  1  8      184.707  0.00   0.235   0.024  KW CHNGE
     3.8312   95.213    6.040  1 14      105.972  0.80   0.385   0.000  KW CHNGE
     3.8312   95.213    6.040  1 14      105.911  0.80   0.385   0.000  CONTACT 
     3.8312   91.732   10.324 14 15        0.000  0.00   0.000  -1.000  COELESCE
 13000.0000   91.732    0.000 14 15        0.000 -1.00   0.000  -1.000  MAX TIME
\end{verbatim}
Note: with $\ftz=0.9$, the final BH mass $=6.040\Ms+(95.213\Ms\times0.9)=91.732\Ms$.
\end{minipage}

\subsection*{Example 2a ($Z=0.0001$, $\ftz=0.0$, $\alpha=3.0$):}
\begin{minipage}[h]{\textwidth}
\begin{verbatim}
      TIME      M1       M2   K1 K2        SEP    ECC  R1/ROL1 R2/ROL2  TYPE
     0.0000  182.918   21.690  1  1     1227.847  0.20   0.017   0.015  INITIAL 
     3.1533  180.026   21.688  2  1     1245.455  0.20   0.023   0.017  KW CHNGE
     3.1540  179.941   21.688  2  1     1245.944  0.20   1.001   0.017  BEG RCHE
     3.1540   84.025   21.688  7  1       17.855  0.00   1.001   0.017  COMENV  
     3.1540   84.025   21.688  7  1       17.855  0.00   0.383   0.973  END RCHE
     3.2029   83.625   21.692  7  1       17.405  0.00   0.392   1.000  BEG RCHE
     3.3449   82.129   21.709  7  1       17.477  0.00   0.387   1.000  END RCHE
     3.3576   81.971   21.710 15  1        0.000  0.00   0.000  -1.000  NO REMNT
     9.0197    0.000   21.699 15  2        0.000 -1.00  -1.000   0.000  KW CHNGE
     9.0359    0.000   21.699 15  4        0.000 -1.00  -1.000   0.000  KW CHNGE
     9.8854    0.000   21.172 15  5        0.000 -1.00  -1.000   0.000  KW CHNGE
     9.8928    0.000   19.036 15 14        0.000 -1.00  -1.000   0.000  KW CHNGE
 13000.0000    0.000   19.036 15 14        0.000 -1.00  -1.000   0.000  MAX TIME
\end{verbatim}
\end{minipage}

\subsection*{Example 2b ($Z=0.0001$, $\ftz=0.0$, $\alpha=1.0$):}
\begin{minipage}[h]{\textwidth}
\begin{verbatim}
      TIME      M1       M2   K1 K2        SEP    ECC  R1/ROL1 R2/ROL2  TYPE
     0.0000  182.918   21.690  1  1     1227.847  0.20   0.017   0.015  INITIAL 
     3.1533  180.026   21.688  2  1     1245.455  0.20   0.023   0.017  KW CHNGE
     3.1540  179.941   21.688  2  1     1245.944  0.20   1.001   0.017  BEG RCHE
     3.1540  174.972   21.688  2 15        6.009  0.00   1.001   0.017  COMENV  
     3.1541  174.952    0.000  4 15        0.000 -1.00   0.000  -1.000  KW CHNGE
     3.4342  132.942    0.000  5 15        0.000 -1.00   0.000  -1.000  KW CHNGE
     3.4430  180.026    0.000 15 15         MAX TIME  0.00   0.000  -2.000  NO REMNT
\end{verbatim}
\end{minipage}

\subsection*{Example 2c ($Z=0.001$, $\ftz=0.0$, $\alpha=3.0$):}
\begin{minipage}[h]{\textwidth}
\begin{verbatim}
      TIME      M1       M2   K1 K2        SEP    ECC  R1/ROL1 R2/ROL2  TYPE
     0.0000  182.918   21.690  1  1     1227.847  0.20   0.019   0.018  INITIAL 
     3.1757  113.436   21.688  2  1     1857.539  0.20   0.038   0.012  KW CHNGE
     3.1769  113.270   21.689  2  1     1859.623  0.20   1.001   0.012  BEG RCHE
     3.1769   53.126   21.689  7  1       34.248  0.00   1.001   0.012  COMENV  
     3.1769   53.126   21.689  7  1       34.248  0.00   0.165   0.524  END RCHE
     3.4405   40.500   21.733 14  1       41.525  0.09   0.000   0.411  KW CHNGE
     3.4405   40.500   21.733 14  1       41.501  0.09   0.000   0.411  BEG SYMB
     8.9310   40.518   21.668 14  1       40.936  0.00   0.000   1.001  BEG RCHE
     9.0679   40.519   21.665 14  1       40.934  0.00   0.000   0.981  END RCHE
     9.0679   40.519   21.665 14  1       40.934  0.00   0.000   0.981  BEG SYMB
     9.0941   40.519   21.665 14  2       40.995  0.00   0.000   0.838  KW CHNGE
     9.0963   40.519   21.664 14  2       41.004  0.00   0.000   1.001  BEG RCHE
     9.1091   40.522   21.471 14  4       41.674  0.00   0.000   2.815  KW CHNGE
     9.6953   40.809    7.693 14  4      203.643  0.00   0.000   0.655  END RCHE
     9.6953   40.809    7.693 14  4      203.643  0.00   0.000   0.655  BEG SYMB
     9.7154   40.815    7.682 14  7      203.593  0.00   0.000   0.018  KW CHNGE
     9.7154   40.815    7.682 14  7      203.593  0.00   0.000   0.018  BEG SYMB
     9.9688   40.816    7.600 14  8      203.931  0.00   0.000   0.016  KW CHNGE
    10.0068   40.816    1.818 14 13       33.618  7.07   0.000  -2.000  DISRUPT 
 13000.0000   40.816    1.818 14 13        0.000 -1.00   0.000   0.000  MAX TIME
\end{verbatim}
\end{minipage}

\subsection*{Example 3a ($Z=0.0001$, $\ftz=0.0$, $\alpha=3.0$):}
\begin{minipage}[h]{\textwidth}
\begin{verbatim}
      TIME      M1       M2   K1 K2        SEP    ECC  R1/ROL1 R2/ROL2  TYPE
     0.0000  198.903  104.554  1  1       63.895  0.50   0.431   0.454  INITIAL 
     2.4086  197.215  104.291  1  1       46.743  0.00   0.759   1.001  BEG RCHE
     3.1312  196.235  103.562  1  1       47.017  0.00   0.729   1.459  BEG BSS 
     3.1349  196.238  103.546  2  1       47.319  0.00   0.618   1.454  KW CHNGE
     3.1349  196.238  103.546  2  1       47.319  0.00   1.002   1.454  CONTACT 
     3.1349  299.784  103.546  2 15        7.965  0.00   0.990   0.990  COMENV  
     3.1353  299.779    0.000  4 15        0.000 -1.00   0.000  -1.000  KW CHNGE
     3.4033  259.580    0.000  5 15        0.000 -1.00   0.000  -1.000  KW CHNGE
     3.4081  232.983    0.000 14 15        0.000 -1.00   0.000  -1.000  KW CHNGE
 13000.0000  232.983    0.000 14 15        0.000 -1.00   0.000  -1.000  MAX TIME
\end{verbatim}
\end{minipage}

\subsection*{Example 3b ($Z=0.02$, $\ftz=0.0$, $\alpha=3.0$, $0.2\times{\rm B10~wind}$):}
\begin{minipage}[h]{\textwidth}
\begin{verbatim}
      TIME      M1       M2   K1 K2        SEP    ECC  R1/ROL1 R2/ROL2  TYPE
     0.0000  191.915  130.275  1  1       56.154  0.00   1.174   1.066  INITIAL 
     0.0000  191.915  130.275  1  1       56.154  0.00   1.174   1.066  BEG RCHE
     0.0000  191.915  130.275  1  1       53.190  0.00   1.240   1.125  CONTACT 
     0.0000  296.135  130.275  1 15        0.000  0.00   0.000  -1.000  COELESCE
     3.0270  205.443    0.000  2 15        0.000 -1.00   0.000  -1.000  KW CHNGE
     3.0291  205.379    0.000  4 15        0.000 -1.00   0.000  -1.000  KW CHNGE
     3.3598  195.457    0.000  5 15        0.000 -1.00   0.000  -1.000  KW CHNGE
     3.3694  175.651    0.000 14 15        0.000 -1.00   0.000  -1.000  KW CHNGE
 13000.0000  175.651    0.000 14 15        0.000 -1.00   0.000  -1.000  MAX TIME
\end{verbatim}
Note: with $\fmrg=0.2$, the mass of the merger product $191.915\Ms+130.275\Ms\times(1-0.2)=296.135\Ms$. 
\end{minipage}

\subsection*{Example 3c ($Z=0.02$, $\ftz=0.0$, $\alpha=3.0$):}
\begin{minipage}[h]{\textwidth}
\begin{verbatim}
      TIME      M1       M2   K1 K2        SEP    ECC  R1/ROL1 R2/ROL2  TYPE
     0.0000  191.915  130.275  1  1       56.154  0.00   1.174   1.066  INITIAL 
     0.0000  191.915  130.275  1  1       56.154  0.00   1.174   1.066  BEG RCHE
     0.0000  191.915  130.275  1  1       53.190  0.00   1.240   1.125  CONTACT 
     0.0000  296.135  130.275  1 15        0.000  0.00   0.000  -1.000  COELESCE
     3.1652   62.867    0.000  2 15        0.000 -1.00   0.000  -1.000  KW CHNGE
     3.1690   62.340    0.000  4 15        0.000 -1.00   0.000  -1.000  KW CHNGE
     3.3890   29.340    0.000  7 15        0.000 -1.00   0.000  -1.000  KW CHNGE
     3.5778   16.768    0.000  8 15        0.000 -1.00   0.000  -1.000  KW CHNGE
     3.5813   14.968    0.000 14 15        0.000 -1.00   0.000  -1.000  KW CHNGE
 13000.0000   14.968    0.000 14 15        0.000 -1.00   0.000  -1.000  MAX TIME
\end{verbatim}
\end{minipage}

\twocolumn

\end{document}